\documentclass[journal]{IEEEtran}
\usepackage{amsmath}
\usepackage{amssymb}
\usepackage{algorithm}
\usepackage{algorithmicx}
\usepackage{algpseudocode}
\usepackage[T1]{fontenc}
\usepackage{color}
\usepackage{times}
\usepackage{cite}

\usepackage{caption}
\usepackage{graphicx}
\usepackage{float} 
\usepackage{subcaption}
\usepackage{setspace}
\usepackage{stfloats}

\begin{document}
	\newcounter{TempEqCnt}
\title{\vspace{-20pt}Exploring Hybrid Active-Passive RIS-Aided MEC Systems: From the Mode-Switching Perspective}
\author{Hao Xie,  ~\IEEEmembership{Graduate Student Member, IEEE,} Dong Li,~\IEEEmembership{Senior Member,~IEEE,} Bowen Gu, ~\IEEEmembership{Graduate Student Member, IEEE,} \vspace{-30pt}
	\thanks{H. Xie, D. Li, and B. Gu are with the School of Computer Science and Engineering, Macau University of Science and Technology, Avenida Wai Long, Taipa, Macau 999078, China (e-mails: 3220005631@student.must.edu.mo, dli@must.edu.mo, 21098538ii30001@student.must.edu.mo).}
} 

\maketitle

\begin{abstract}
Mobile edge computing (MEC) has been regarded as a promising technique to support latency-sensitivity and computation-intensive serves. However, the low offloading rate caused by the random channel fading characteristic becomes a major bottleneck in restricting the performance of the MEC. Fortunately, reconfigurable intelligent surface (RIS) can alleviate this problem since it can boost both the spectrum- and energy- efficiency. Different from the existing works adopting either fully active or fully passive RIS, we propose a novel hybrid RIS in which reflecting units can flexibly switch between active and passive modes. To achieve a tradeoff between the latency and energy consumption, an optimization problem is formulated by minimizing the total cost via jointly optimizing the transmission time, the transmit power, the receive beamforming vector, the phase-shift matrix, the mode-switching factor, the amplification factor, the offloading ratio factor, and the computation ability of the user, where the constraints of the maximum energy of users, the maximum power of the RIS, the minimum computation tasks, the transmission time, the offloading ratio factor, the mode-switching factor, the unit moduli of passive units, and the computation ability are taken into account. Considering the complexity of the aforementioned problem, we develop an alternating optimization-based iterative algorithm by combining the successive convex approximation method, the variable substitution, and the singular value decomposition to obtain sub-optimal solutions. Furthermore, in order to gain more insight into the problem, we consider two special cases involving a latency minimization problem and an energy consumption minimization problem, and respectively analyze the tradeoff between the number of active and passive units. Simulation results verify that the proposed algorithm can achieve flexible mode switching and significantly outperforms existing algorithms.
\end{abstract}


\vspace{-22pt}
\section{Introduction}\vspace{-5pt}
\subsection{Backgroud}
The development and popularization of Internet of Things technologies have greatly facilitated the realization of intelligent applications (e.g., augmented reality, autonomous navigation, speech recognition, e-Health care, etc.)\cite{a1}. However, these applications and services generally require a massive number of devices to perform computation-intensive and latency-sensitive tasks, which poses tremendous challenges for size-constrained and low-power devices. Mobile cloud computing, which leverages computing and storage resources from powerful centralized clouds located remotely, has been considered an effective method for offloading computation tasks\cite{a2}. However, although mobile cloud computing can enhance the computing capacity of the user, there are still some potential problems with its application, such as intolerably high latency and unstable connections caused by long-distance transmission and backhaul.

To supplement mobile cloud computing and address these potential problems, mobile edge computing (MEC), an extension of traditional mobile cloud computing, has emerged as a promising solution\cite{b1}. Different from mobile cloud computing, the distributed MEC server is deployed on the edge of wireless networks and is closer to users, which can provide lower latency and save backhaul bandwidth. Generally, there exist two offloading modes for the MEC\cite{b2}: 1) binary offloading, which is suitable for simple tasks that are not partitioned; 2) partial offloading, which is suitable for complex tasks composed of multiple parallel segments. However, there are still some potential problems with the offloading process. The offloading rate is typically low when the channel condition is at a poor level, which will lead to a high latency and an even worse performance than that of fully local computing. Thus, many research efforts have been paid to investigate how to improve the offloading rate of the MEC with performance guarantee\cite{b3,b4,b5,b6,b7,b8}. Generally, there exist four typical techniques to solve these problems\cite{b9}. The first technology is to form heterogeneous networks by deploying massive small base stations, which can enhance access availability and spectrum utilization\cite{b3}. This technology allows users to select other edge servers for cooperative offloading, and the tasks that cannot be processed by the MEC servers are further offloaded to a cloud center, which mitigates the computation burden\cite{b4}. The second technology is to deploy massive antennas at the base station to enhance the array and diversity gain, and interference suppression capacity \cite{b5}. This technology can effectively improve the spectrum efficiency without increasing the transmit power and bandwidth, which has become one of the crucial technologies for 5G and its beyond\cite{b6}. The third technology extends the available bandwidth to higher frequency bands such as millimeter-wave\cite{b7}. Millimeter-wave technology is suitable for MEC-aided systems due to its substantial bandwidth and high transmission rate, which can exponentially increase the offloading ability of the MEC. The fourth technology is to deploy the MEC functionality to the unmanned aerial vehicle (UAV), and the UAV can exploit its mobility, flexibility, and maneuverability to maintain the line-of-sight transmission of information and improve the offloading rate, especially when the communication infrastructures are destroyed by natural disasters\cite{b8}. Despite their effectiveness in improving the performance of the MEC, the above technologies also suffer from the following problems including high deployment costs, complex hardware structure, high energy costs, and complex signal processing. Thus, the above observations motivate us to seek other effective solutions to the MEC in the era of 6G.

Recently, reconfigurable intelligent surface (RIS), also known as intelligent reflecting surface (IRS), has received much attention from academia and industry and has been well recognized as a potential technology for 6G\cite{c1,c1-1}. Due to its advantages, the RIS-aided wireless communication has attracted significant attention in channel estimation \cite{c2,c3}. Moreover, the RIS  also promotes the development of various advanced technologies, such as holographic multiple-input multiple-output surface, which can incorporate densely packed sub-wavelength patch antennas to achieve programmable wireless environments\cite{c4}. Compared with the four typical techniques mentioned above, the RIS can enhance both the spectral- and energy- efficiency but with low energy consumption and low cost, which paves a new way for the stable transmission of the MEC-aided system. So far, there have been already some works investigating RIS-aided communication systems\cite{d1,d2,d2-1,d3,d3-1,d2-2,d4,d4-1}. Especially, latency minimization problems were investigated in \cite{d1} for single-user and multi-user scenarios, which verified that the RIS could significantly reduce the latency compared to traditional MEC-aided systems. Furthermore, RIS-aided MEC systems were extended into wireless-powered communication networks to provide a sustainable energy supply and minimize energy consumption \cite{d2}. The energy consumption was optimized in \cite{d2-1} by designing phase shift, data size, transmission rate, power control, and decoding order, demonstrating the proposed algorithm's high energy efficiency. Furthermore, the authors in \cite{d2-2} explored optimization-based and data-driven solutions to maximize total completed task-input bits, proposed a three-step block coordinate descending algorithm for efficient solutions, and constructed deep learning architectures for online implementation, showing promising performance and practicality. The authors in \cite{d3} evaluated the impact of the RIS on the computational performance, where partial computational offloading was considered. Different from previous works, the work \cite{d3-1} compared the performance of time-division multiple access (TDMA) and non-orthogonal multiple access (NOMA) schemes for uplink offloading, employing three different RIS beamforming schemes to strike a balance between the system performance and the signaling overhead.
To obtain the optimal tradeoff between achievable rate and energy consumption, the computational energy efficiency maximization was considered in \cite{d4} via the NOMA. The work \cite{d4-1} extended the passive RIS to the active RIS, investigating the performance of the active RIS under different multiple access schemes. Besides, it proposed a hybrid TDMA-NOMA scheme and validates the performance of the active RIS.
\vspace{-10pt}
\subsection{Motivation and Contributions}
Despite the above works, some fundamental issues still remain unsolved in RIS-aided MEC systems. \textit{\textbf{On one hand}, the passive RIS will suffer from the ``doubly path loss'' effect, i.e., the path loss of the cascaded channel is much larger than that of the direct channel, which has become a significant bottleneck in restricting the performance of the RIS-aided MEC system. Although there exist some schemes to deal with this problem, such as deploying a large number of reflecting units to achieve higher passive beamforming gains, these schemes will increase significant phase shift feedback compression overhead \cite{e1} in the feedback channel and thus may be unaffordable in practice. \textbf{On the other hand}, the active RIS can not only adjust the phase shift to achieve the passive beamforming as passive RIS but also amplify the received signal, which can indeed mitigate the ``doubly path loss'' effect\cite{e2}. However, in view of the energy consumption on the RIS side, it is unreasonable to always adopt the fully-active architecture for the RIS since the power consumption of the active RIS, which is much higher than that of the passive RIS, cannot be ignored. Besides, there exist some works that have shown that the advantages between the active RIS and the passive RIS are complementary\cite{e3,e4}, i.e., the active RIS does not always outperform the passive RIS. Thus, neither fully active RIS nor fully passive RIS can effectively achieve the full potential of the RIS.}

Motivated by the above observations, we investigate and analyze a hybrid RIS-aided MEC system\footnote{The hardware complexity of the hybrid RIS is nearly identical to that of the active RIS since we only need to control whether access to the reflection-type amplifiers on the basis of the active RIS, the mode-switching mechanism can be realized.}, where the reflecting units in the hybrid RIS can be switched into two modes flexibly\footnote{When a large number of reflecting units are employed, we can utilize phase-shift compression techniques \cite{e1} to significantly reduce the signaling overhead.}, i.e., passive mode and active mode. Passive units can achieve passive beamforming by adjusting the phase shift when the reflecting units are switched into the passive mode, while active units can not only adjust the phase shift but also amplify the received signal when reflecting units are switched into the active mode. By applying the mode-switching scheme, the MEC system can achieve performance complementarity between the passive and active RISs, compensating for their respective shortcomings and achieving low latency while ensuring low energy consumption. This allows the MEC system to better adapt to various communication scenarios and application requirements. Note that the hybrid RIS-aided systems have received little attention, and there are only \cite{f1,f2} related to this work regarding the hybrid RIS. However, the joint optimization of one active RIS and one passive RIS was considered in \cite{f1}. The optimization of the number of active and passive units in the hybrid RIS was explored in \cite{f2} under a given power budget, which is different from the proposed mode-switching scheme under a given number of reflecting units in our work. The main contributions of the paper are summarized as follows.
\begin{itemize}
	\item For the hybrid RIS-aided MEC system, each user divides its computational tasks into two parts based on partial offloading. One part is computed locally, and the other part is offloaded to the MEC with the help of the hybrid RIS via a TDMA protocol. To achieve a tradeoff between the latency and energy consumption, the total cost is minimized by jointly optimizing the transmission time, the transmit power, the receive beamforming vector, the phase-shift matrix, the mode-switching factor, the amplification factor, the offloading ratio factor, and the computation ability subject to the maximum energy constraints of users, the maximum power constraint of the RIS, the minimum computation tasks constraint, the transmission time constraint, the offloading ratio factor constraint, the mode-switching factor constraint, the unit-modulus constraints of passive units, and the computation ability constraint.
	\item In light of the intractability of the formulated problem, we develop an alternating optimization (AO)-based algorithm that adopts the variable substitution method, the successive convex approximation (SCA) method, and the singular value decomposition (SVD) method to obtain the corresponding sub-optimal solutions. In particular, the closed-form solution of the receive beamforming vector is obtained based on linear minimum-mean-square-error (MMSE) detection. 
	\item To gain more insight into the tradeoff between the number of active and passive reflecting units, we reduce the original optimization problem to a latency minimization problem and an energy consumption minimization problem, respectively. In particular, for the latency minimization problem, we derive the closed-form solutions of the amplification factor and the number of active units, which shows all the reflecting units will be switched to the active mode when the channel condition is poor, while the reflecting units will be gradually switched to the passive mode when the channel condition is better than a certain value. For the energy consumption minimization problem, the offloading requirement must be guaranteed. The offloading outage will occur when the offloading task is large, or the transmit power is low and the channel condition is poor. Then, increasing the number of active units and amplification factor can effectively improve the offloading rate to guarantee offloading requirements.	
	\item Simulation results verify that the effectiveness of the proposed mode-switching algorithm and the proposed algorithm outperforms baseline algorithms. Moreover, the reflecting units under the proposed algorithm can flexibly switch between active and passive modes to minimize the total cost.	
\end{itemize}

\textit{Notations:} In this paper, scalars are represented by lowercase letters, vectors are represented by boldface lowercase letters, and matrices are represented by boldface uppercase letters. 
$|\cdot|$ represents the absolute value and $\|\cdot\|$ signifies the Euclidean norm. $\mathbb{E}(\cdot)$ stands for the statistical expectation, while $\mathbb{C}^{N \times M}$ represents the $N\times M$ complex-valued matrix. $\mathcal{CN}(\mu,\sigma^2)$ represents the distribution of a circularly symmetric complex Gaussian random variable with a mean of $\mu$ and a variance of $\sigma^2$.  ${\rm diag}(\cdot)$, ${\rm Tr}(\cdot)$, $\boldsymbol{\rm X}^T$, $\boldsymbol{\rm X}^H$, and $[\boldsymbol{{\rm X}}]_{i,j}$ represent the diagonalization, the trace, the transpose, the conjugate transpose, and the $(i,j)$-th element matrix $\boldsymbol{\rm X}$,  respectively. $\boldsymbol{{\rm I}}$ and $\boldsymbol{0}$ denote the identity matrix and all-zero matrix, respectively. $\boldsymbol{\rm X}\succeq \boldsymbol{0}$ means that $\boldsymbol{\rm X}$ is a positive semidefinite matrix. $\arg(\cdot)$ denotes the angle of a complex number.
\subsection{Organization}
The following structure is applied in this paper: Section  \uppercase\expandafter{\romannumeral2} describes the system model and problem formulation. Section \uppercase\expandafter{\romannumeral3} presents the mode-switching algorithm for hybrid RIS. Section \uppercase\expandafter{\romannumeral4} provides the tradeoff between the number of active and passive reflecting units. Section \uppercase\expandafter{\romannumeral5} shows the simulation results, and this paper is concluded in Section \uppercase\expandafter{\romannumeral6}.
\section{System Model and Problem Formulation}
\begin{figure}[!t]
	\centering
	\includegraphics[width=2.5in]{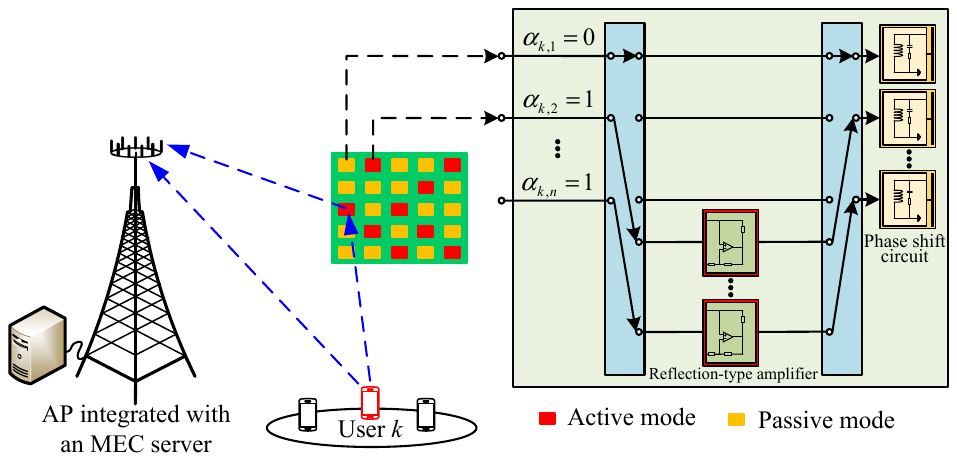}
	\caption{A hybrid RIS-aided MEC system.}
	\label{fig0}\vspace{-10pt}
\end{figure}

\begin{figure}[!t]
	\centering
	\includegraphics[width=1.5in]{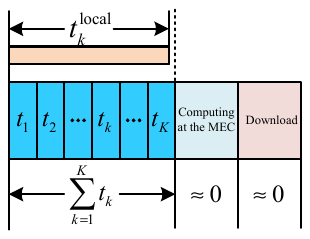} 
	\caption{The transmission time slot structure.}
	\label{fig0-1}
\end{figure}

As illustrated in Fig. \ref{fig0}, we consider a hybrid RIS-aided MEC system consisting of an access point (AP) with $M$ antennas, an MEC, a hybrid RIS, and $K$ single-antenna users. Users offload tasks to the MEC via a TDMA protocol\footnote{Both TDMA \cite{f2-a} and spatial division multiple access (SDMA) schemes  are feasible solutions for task offloading. Adopting the SDMA scheme for task offloading can enhance the offloading efficiency since it strikes a balance between the spatial diversity and the spatial multiplexing \cite{f2-b}. The mode switching in the SDMA manner is interesting to be investigated, which is, however,  left to the future work due to space limitations.}. The latency between the AP and the MEC is deemed to be negligible since the MEC and the AP are assumed to be co-located and connected by high-throughput and low-latency optical fiber. The hybrid RIS equipped with $N$ reflecting units is deployed in the cell for assisting the computation offloading of users, where reflecting units can switch modes between active and passive modes. The quasi-static flat-fading channel is assumed in this system, where the channel state information (CSI) remains constant within a channel coherence frame but may change for different frames\footnote{Under the quasi-static flat fading channel, the proposed algorithm only needs to be executed at the beginning of the frames and does not need frequent updates, and thus significantly reducing the signaling overhead\cite{f2-1}. Furthermore, we can leverage the method presented in \cite{f2} to extend the design with instantaneous CSI to that with statistical CSI or apply signaling compression techniques \cite{f3} to reduce the signaling overhead.}. The transmission time slot structure of the considered system is shown in Fig. \ref{fig0-1}. Here, the computing time and downloading time can be ignored since the MEC has a much stronger computing capability than users, and the number of the bits related to the computing result is very small, which is commonly utilized in existing works (see, e.g., \cite{h0,h1}). Meanwhile, each user can perform local computing within the total offloading time since each user has a separated circuit architecture between the computing unit and the offloading unit\cite{h1,h1a}.

We assume that each user has a computation task and denote $T_k = (S_k,C_k)$ as the task for user $k$, where $S_k$ is the input data size, and its unit is bits. $C_k$ denotes the total number of CPU cycles
required to execute this task, and its unit is cycles.  Consider a partial offloading strategy, where each user processes its computation tasks via local computing and task offloading. In what follows, we give the local computing process and task offloading process.
\vspace{-10pt}
\subsection{Local Computing}
Define $f_k^{\textrm{local}}$ as the computation ability for user $k$, the time required for local computing is given by\cite{b2,h1b,h1c}
\begin{equation}\label{eq1}
\begin{array}{l}
t_k^{\textrm{local}}=\frac{(1-\beta_k)C_k}{f_k^{\textrm{local}}},
\end{array}
\end{equation}
where $\beta_k$ denotes the offloading ratio factor. The energy consumption of the $k$-th user for local computing can be expressed as\cite{b2,h1b,h1c}
\begin{equation}\label{eq2}
\begin{array}{l}
E_k^{\textrm{local}}=(1-\beta_k)C_k\kappa {(f_k^{\textrm{local}})}^2,
\end{array}
\end{equation}
where $\kappa$ is a coefficient depending on the chip architecture.
\subsection{Task Offloading}
The received signal at the AP is formulated as
\begin{equation}\label{eq3}
\begin{array}{l}
\boldsymbol{\rm y }_k
=(\boldsymbol{\rm h}_{{\rm d}, k}+\boldsymbol{\rm H}^H\boldsymbol{\rm \Lambda}_k\boldsymbol{\rm \Theta}_k\boldsymbol{\rm h}_{{\rm r}, k})\sqrt{p_k}s_k\\
~~~~~~~~~~~~~~~~~~~~~~~~~~+\boldsymbol{\rm H}^H\boldsymbol{\rm A}_k\boldsymbol{\rm \Lambda}_k\boldsymbol{\rm \Theta}_k\boldsymbol{\rm z}_k+\boldsymbol{\rm n}_k,
\end{array}
\end{equation}
where $p_k$ and $s_k$ denote the transmit power and the offloading signal of the $k$-th user, respectively. $\boldsymbol{\rm h}_{{\rm d},k}\in\mathbb{C}^{M\times 1}$ and $\boldsymbol{\rm h}_{{\rm r},k}\in\mathbb{C}^{N\times 1}$ denote the channel vectors from the $k$-th user to the AP and the RIS, respectively. $\boldsymbol{\rm H}\in\mathbb{C}^{N\times M}$ denotes the channel matrix from the RIS to the AP. $\boldsymbol{\rm \Theta}_k\triangleq{\textrm{diag}}(e^{j\theta_{k,1}},\cdots,e^{j\theta_{k,n}},\cdots,e^{j\theta_{k,N}})$ is the diagonal phase-shift matrix in time slot $k$, where $\theta_{k,n}$ denotes the corresponding phase shift. $\boldsymbol{\rm A}_k={\textrm{diag}}(\alpha_{k,1},\cdots,\alpha_{k,N})$ denotes the mode-switching matrix. $\boldsymbol{\rm \Lambda}_k={\textrm{diag}}(\rho_{k,1}^{\alpha_{k,1}},\cdots,\rho_{k,n}^{\alpha_{k,n}},\cdots,\rho_{k,N}^{\alpha_{k,N}})$ denotes the reflecting amplification matrix, where $\rho_{k,n}^{\alpha_{k,n}}$ denotes the amplification factor of the $n$-th reflecting unit. The active mode of the $n$-th reflecting unit is activated, and the amplification factor is $\rho_{k,n}$ when $\alpha_{k,n}=1$; otherwise, the passive mode is activated, and the amplification factor is $1$. $\boldsymbol{\rm z}_k\in\mathbb{C}^{N\times 1}$ is the thermal noise introduced by the active reflecting units due to signal amplification, which is assumed to follow the independent circularly symmetric complex Gaussian distribution, i.e., $\boldsymbol{\rm z}_k\sim\mathcal{CN}(\boldsymbol{0}, \sigma^2\boldsymbol{\rm I}_N)$. $\boldsymbol{\rm n}_k\sim\mathcal{CN}(0,\delta^2\boldsymbol{\rm I}_M)$ denotes the additive white Gaussian noise (AWGN) at the AP.
To enhance the signal offloaded by the $k$-th user, a beamforming vector $\boldsymbol{\rm w}_k\in\mathbb{C}^{M\times 1}$ is applied with $\|\boldsymbol{\rm w}_k\|^2=1$. Thus, the signal recovered at the AP is given by
\begin{equation}\label{eq4}
\begin{array}{l}
y_k=\boldsymbol{\rm w}_k^H\boldsymbol{\rm y }_k\\
~~~=\boldsymbol{\rm w}_k^H(\boldsymbol{\rm h}_{{\rm d}, k}+\boldsymbol{\rm H}^H\boldsymbol{\rm \Lambda}_k\boldsymbol{\rm \Theta}_k\boldsymbol{\rm h}_{{\rm r}, k})\sqrt{p_k}s_k\\
~~~+\boldsymbol{\rm w}_k^H\boldsymbol{\rm H}^H\boldsymbol{\rm A}_k\boldsymbol{\rm \Lambda}_k\boldsymbol{\rm \Theta}_k\boldsymbol{\rm z}_k+\boldsymbol{\rm w}_k^H\boldsymbol{\rm n}_k.
\end{array}
\end{equation}
Then, the signal-to-interference-plus-noise ratio (SINR) of  the $k$-th user’s signal recovered is given by
\begin{equation}\label{eq5}
\begin{array}{l}
\gamma_k=\frac{p_k|\boldsymbol{\rm w}_k^H(\boldsymbol{\rm h}_{{\rm d},k}+\boldsymbol{\rm H}^H\boldsymbol{\rm \Lambda}_k\boldsymbol{\rm \Theta}_k\boldsymbol{\rm h}_{{\rm r}, k})|^2}{\sigma^2\|\boldsymbol{\rm w}_k^H\boldsymbol{\rm H}^H\boldsymbol{\rm A}_k\boldsymbol{\rm \Lambda}_k\boldsymbol{\rm \Theta}_k\|^2+\delta^2}.
\end{array}
\end{equation}
Denote the system bandwidth by $B$ Hz, the achievable computation offloading rate of the $k$-th user is formulated as
\begin{equation}\label{eq6}
\begin{array}{l}
R_k=B\log_2(1+\gamma_k).
\end{array}
\end{equation}
Define $t_k$ as the transmission time of $k$-th user in the task offloading process, and we have the following constraint
\begin{equation}\label{eq7}
\begin{array}{l}
t_kR_k\geq \beta_kS_k.
\end{array}
\end{equation}

The power consumption at the active units and the passive units can be respectively expressed as\cite{e2}
\begin{equation}\label{eq8}
\begin{array}{l}
P_k^{\rm act}=\sum\limits_{n=1}^N\alpha_{k,n}(P_{\textrm{C}}+P_{\textrm{DC}})\\
~~~~~~+p_k\|\boldsymbol{\rm A}_k\boldsymbol{\rm \Lambda}_k\boldsymbol{\rm \Theta}_k\boldsymbol{\rm h}_{{\rm r}, k}\|^2{+}\sigma^2\|\boldsymbol{\rm A}_k\boldsymbol{\rm \Lambda}_k\boldsymbol{\rm \Theta}_k\|^2,\\
\end{array}
\end{equation}
\begin{equation}\label{eq9}
\begin{array}{l}
P_k^{\rm pas}=(N-\sum\limits_{n=1}^N\alpha_{k,n})P_{\textrm{C}},
\end{array}
\end{equation}
where $P_{\textrm{C}}$ and $P_{\textrm{DC}}$ denote the power consumption of the circuit and the DC biasing power consumption, respectively. The total energy consumption of the system is given by
\begin{equation}\label{eq10}
\begin{array}{l}
E^{\rm total}_k=t_kp_k+E_k^{\textrm{local}}+t_kP_k^{\rm act}+t_kP_k^{\rm pas}.
\end{array}
\end{equation}

The active reflecting units need additional power consumption to support the active load, thus they are allowed to allocate the remaining power to amplify the incident signal after supplying the hardware power consumption for $\sum_{n=1}^N\alpha_{k,n}$ active reflecting units. Moreover, the amplification power is limited due to the overall power budget for the active units\cite{e3}. We denote the maximum amplification power of the active units as $P_{\rm R}^{\max}$, and thus we have
\begin{equation}\label{eq11}
\begin{array}{l}
 p_k\|\boldsymbol{\rm A}_k\boldsymbol{\rm \Lambda}_k\boldsymbol{\rm \Theta}_k\boldsymbol{\rm h}_{{\rm r}, k}\|^2+\sigma^2\|\boldsymbol{\rm A}_k\boldsymbol{\rm \Lambda}_k\boldsymbol{\rm \Theta}_k\|^2\leq P_{\rm R}^{\max}.
\end{array}
\end{equation}
\subsection{Problem Formulation}
In this paper, we aim to address the tradeoff between the energy consumption and the latency for the partial task offloading in hybrid RIS-aided MEC systems. Mathematically, the problem is expressed as follows\footnote{The proposed algorithm can achieve dynamic rate adaptation for different frames, but not within a single frame, which, however, is a reasonable scenario since each user does not need to perform task offloading frequently\cite{h1}.}
\begin{equation}\label{eq12}
\begin{array}{l}
\min\limits_{\mbox{\scriptsize$\begin{array}{c} 
		\boldsymbol{\rm \Theta}_k,t_k,p_k\\\beta_k,
		\boldsymbol{\rm w}_k,
		\alpha_{k,n}\\\boldsymbol{\rm \Lambda}_k,f_k^{\textrm{local}}
		\end{array}$}} 
\sum\limits_{k=1}^K w_kt_k+\sum\limits_{k=1}^K(1-w_k)\psi E^{\rm total}_k\\
s.t.~{C_1}: t_kp_k+E_k^{\textrm{local}}\leq E_k^{\max},\\
~~~~~{C_2}: p_k\|\boldsymbol{\rm A}_k\boldsymbol{\rm \Lambda}_k\boldsymbol{\rm \Theta}_k\boldsymbol{\rm h}_{{\rm r}, k}\|^2{+}\sigma^2\|\boldsymbol{\rm A}_k\boldsymbol{\rm \Lambda}_k\boldsymbol{\rm \Theta}_k\|^2{\leq} P_{\rm R}^{\max},\\
~~~~~{C_3}:0\leq \beta_k\leq 1,\\
~~~~~{C_4}:t_kR_k\geq \beta_k S_k,\\
~~~~~{C_5}:\alpha_{k,n}\in\{0,1\},\\
~~~~~{C_6}:\frac{(1-\beta_k)C_k}{f_k^{\textrm{local}}}\leq\sum\limits_{k=1}^Kt_k, t_k\leq T_k^{\max},\\
~~~~~{C_7}:0\leq f_k^{\textrm{local}}\leq f_k^{\max},\\
~~~~~C_8:|[(\boldsymbol{\rm I}-\boldsymbol{\rm A}_k)\boldsymbol{\rm \Lambda}_k\boldsymbol{\rm \Theta}_k]_{n,n}|=[\boldsymbol{\rm I}-\boldsymbol{\rm A}_k]_{n,n},
\end{array}
\end{equation}
where $\psi$ is the normalizing factor and $w_k$ is the tradeoff factor.  $\psi$ is introduced to implement the unitless combination of energy consumption and latency. The normalizing factor can be defined as the ratio of the average latency to the average energy consumption of the system. $E_k^{\max}$ denotes the maximum allowable energy consumed by each user; $f_k^{\max}$ denotes the maximum CPU frequencies for user $k$. $C_1$ and $C_2$ state that energy consumed by user $k$ and the maximum power of the active units should not exceed $E_k^{\max}$ and $P_{\rm R}^{\max}$, respectively; $C_3$ constrains offloading ratio factor; $C_4$ guarantees the minimum required computation task bits of each user; $C_5$ is mode-switching matrix constraint;  $C_6$ states that the local computation task bits at each user should be executed within $\sum_{k=1}^Kt_k$ and the maximum offloading time is not larger than $T_k^{\max}$; $C_7$ constrains the maximum CPU frequencies of each user; $C_8$ is the unit-modulus constraints of passive units. Note that problem (\ref{eq12}) is a highly non-convex optimization problem and challenging to obtain optimal solutions. Specifically, the coupled variables in the objective function, $C_1$, and $C_4$ render problem (\ref{eq12}) challenging to solve. Besides, compared to the traditional passive RIS-aided communication, although the active units avoid the unit-modulus constraint, it also introduces the additional non-convex constraint $C_2$, aggravating the coupling between the optimization variables. Furthermore, compared to the existing works adopting either fully active or fully passive RIS, the binary variable introduced by the mode-switching mechanism also renders the problem to be a mixed-integer problem. To handle the non-convex problem (\ref{eq12}), we first apply the MMSE principle to obtain the closed-form solutions of the receive beamforming vector. Second, based on the variable substitution, we successfully decouple the transmit power and the transmission time. Third, based on the SCA technique, we relax the binary variable as a continuous variable. Finally, we adopt the Big-M formulation and the SVD method to decouple the phase shift and the amplification factor.

\section{The Mode-switching Algorithm For The Hybrid RIS}
For any given $p_k$, $\boldsymbol{\rm A}_k$, $\boldsymbol{\rm \Lambda}_k$, and $\boldsymbol{\rm \Theta}_k$, it is well-known that the linear MMSE detector is the optimal receive beamforming. Thus, the MMSE-based receive beamforming is written in (\ref{eq13}). 
\begin{figure*}[hb]
	\vspace{-10pt}
		\hrulefill
	\begin{equation}\label{eq13}
	\begin{array}{l}
	\boldsymbol{\rm w}_k{=}\frac{[(\boldsymbol{\rm h}_{{\rm d},k}{+} \boldsymbol{\rm H}^H\boldsymbol{\rm \Lambda}_k\boldsymbol{\rm \Theta}_k\boldsymbol{\rm h}_{{\rm r},k} )(\boldsymbol{\rm h}_{{\rm d},k}{+} \boldsymbol{\rm H}^H\boldsymbol{\rm \Lambda}_k\boldsymbol{\rm \Theta}_k\boldsymbol{\rm h}_{{\rm r},k} )^H{+}\frac{\sigma^2}{p_k}\boldsymbol{\rm H}^H\boldsymbol{\rm A}_k\boldsymbol{\rm \Lambda}_k\boldsymbol{\rm \Theta}_k\boldsymbol{\rm \Theta}_k^H\boldsymbol{\rm \Lambda}_k^H\boldsymbol{\rm A}_k^H\boldsymbol{\rm H}{+}\frac{\delta^2}{p_k}\boldsymbol{\rm I}_M]^{-1}(\boldsymbol{\rm h}_{{\rm d},k}{+} \boldsymbol{\rm H}^H\boldsymbol{\rm \Lambda}_k\boldsymbol{\rm \Theta}_k\boldsymbol{\rm h}_{{\rm r},k} )}{\|[(\boldsymbol{\rm h}_{{\rm d},k}{+} \boldsymbol{\rm H}^H\boldsymbol{\rm \Lambda}_k\boldsymbol{\rm \Theta}_k\boldsymbol{\rm h}_{{\rm r},k} )(\boldsymbol{\rm h}_{{\rm d},k}{+} \boldsymbol{\rm H}^H\boldsymbol{\rm \Lambda}_k\boldsymbol{\rm \Theta}_k\boldsymbol{\rm h}_{{\rm r},k} )^H{+}\frac{\sigma^2}{p_k}\boldsymbol{\rm H}^H\boldsymbol{\rm A}_k\boldsymbol{\rm \Lambda}_k\boldsymbol{\rm \Theta}_k\boldsymbol{\rm \Theta}_k^H\boldsymbol{\rm \Lambda}_k^H\boldsymbol{\rm A}_k^H\boldsymbol{\rm H}{+}\frac{\delta^2}{p_k}\boldsymbol{\rm I}_M]^{-1}(\boldsymbol{\rm h}_{{\rm d},k}{+} \boldsymbol{\rm H}^H\boldsymbol{\rm \Lambda}_k\boldsymbol{\rm \Theta}_k\boldsymbol{\rm h}_{{\rm r},k} )\|},
	\end{array}
	\end{equation}\vspace{-20pt}
\end{figure*}
\subsection{Optimizing $p_k$, $t_k$, and $\beta_k$}
For given $\boldsymbol{\rm w}_k$, $\boldsymbol{\rm A}_k$, $\boldsymbol{\rm \Lambda}_k$, $\boldsymbol{\rm \Theta}_k$, and $f_k^{\rm local}$, problem (\ref{eq12}) is reduced to
\begin{equation}\label{eq14}
\begin{array}{l}
\min\limits_{\mbox{\scriptsize$\begin{array}{c} 
		t_k,p_k,\beta_k\\
		\end{array}$}} 
\sum\limits_{k=1}^K w_kt_k+\sum\limits_{k=1}^K(1-w_k)\psi E^{\rm total}_k\\
~~~~~~~~s.t.~{C_1}-{C_4},{C_6}.
\end{array}
\end{equation}
It is noted that $p_k$ and $t_k$ are coupled with each other. To this end, we define $\bar p_k = p_kt_k$, then $E^{\rm total}_k$, $C_2$, and $C_4$ can be transformed into
\begin{equation}\label{eq15}
\begin{array}{l}
\!\!\!\bar E^{\rm total}_k=\bar p_k+E_k^{\textrm{local}}+t_k\sum\limits_{n=1}^N\alpha_{k,n}(P_{\textrm{C}}+P_{\textrm{DC}})\\
\!\!\!{+}(\bar p_k\|\boldsymbol{\rm A}_k\boldsymbol{\rm \Lambda}_k\boldsymbol{\rm \Theta}_k\boldsymbol{\rm h}_{{\rm r}, k}\|^2{+}t_k\sigma^2\|\boldsymbol{\rm A}_k\boldsymbol{\rm \Lambda}_k\boldsymbol{\rm \Theta}_k\|^2){+}t_kP_k^{\rm pas},
\end{array}
\end{equation}
\begin{equation}\label{eq16}
\begin{array}{l}
\!\!\!\!\!\!\!{\bar C_2}{:} \bar p_k\|\boldsymbol{\rm A}_k\boldsymbol{\rm \Lambda}_k\boldsymbol{\rm \Theta}_k\boldsymbol{\rm h}_{{\rm r}, k}\|^2{+}t_k\sigma^2\|\boldsymbol{\rm A}_k\boldsymbol{\rm \Lambda}_k\boldsymbol{\rm \Theta}_k\|^2{\leq} t_kP_{\rm R}^{\max},\\\end{array}
\end{equation}
\begin{equation}\label{eq17}
\begin{array}{l}
\!\!\!\!\!\!\bar C_4:t_kB\log_2(1{+}\frac{\bar p_k|\boldsymbol{\rm w}_k^H(\boldsymbol{\rm h}_{{\rm d},k}{+}\boldsymbol{\rm H}^H\boldsymbol{\rm \Lambda}_k\boldsymbol{\rm \Theta}_k\boldsymbol{\rm h}_{{\rm r}, k})|^2}{t_k(\sigma^2\|\boldsymbol{\rm w}_k^H\boldsymbol{\rm H}^H\boldsymbol{\rm A}_k\boldsymbol{\rm \Lambda}_k\boldsymbol{\rm \Theta}_k\|^2{+}\delta^2)}){\geq} \beta_k S_k.
\end{array}
\end{equation}
Then, optimization problem (\ref{eq14}) can be reformulated as the following problem
\begin{equation}\label{eq18}
\begin{array}{l}
\min\limits_{\mbox{\scriptsize$\begin{array}{c} 
		t_k,\bar p_k,\beta_k\\
		\end{array}$}} 
\sum\limits_{k=1}^K w_kt_k+\sum\limits_{k=1}^K(1-w_k)\psi \bar E^{\rm total}_k\\
~~~~~~~s.t.~{\bar C_2},{C_3},{\bar C_4},{C_6},{\bar C_1}:~ \bar  p_k+E_k^{\textrm{local}}\leq E_k^{\max}.\\
\end{array}
\end{equation}
Problem (\ref{eq18}) is a convex optimization problem and can be solved by standard convex optimization techniques.
\vspace{-10pt}
\subsection{Optimizing $\boldsymbol{\rm \Theta}_k$, $\alpha_{k,n}$, $\boldsymbol{\rm \Lambda}_k$, and $f_k^{\textrm{local}}$}
For given $p_k$, $t_k$, and $\beta_k$, we optimize $\boldsymbol{\rm \Theta}_k$, $\alpha_{k,n}$, $\boldsymbol{\rm \Lambda}_k$, and $f_k^{\textrm{local}}$, problem (\ref{eq12}) can be reduced to
\begin{equation}\label{eq19}
\begin{array}{l}
\min\limits_{\mbox{\scriptsize$\begin{array}{c} 
		\boldsymbol{\rm \Theta}_k,
		\boldsymbol{\rm w}_k,
		\alpha_{k,n}\\\boldsymbol{\rm \Lambda}_k,f_k^{\textrm{local}}
		\end{array}$}} 
\sum\limits_{k=1}^K w_kt_k+\sum\limits_{k=1}^K(1-w_k)\psi E^{\rm total}_k\\
~~~~~~~~~~~~s.t.~{C_1},{C_2},{C_4}-{C_8}.
\end{array}
\end{equation}
\subsubsection{The Transformation of $C_5$}
For the binary variable $\alpha_{k,n}$, we first relax it to the continuous real variable in the range of [0, 1] by the convex relaxation method, then impose a constraint in the opposite direction to make $\alpha_{k,n}\leq 0$ and $\alpha_{k,n}\geq 1$, i.e., $\alpha_{k,n}-\alpha_{k,n}^2\leq 0$. By applying these constraints, the value of $\alpha_{k,n}$ is well optimized to either 0 or 1. Thus, we equivalently transform $C_5$ as\cite{h2}
\begin{equation}\label{eq20}
\begin{array}{l}
C_{5a}:\alpha_{k,n}-\alpha_{k,n}^2\leq 0,~~~ C_{5b}:0\leq \alpha_{k,n}\leq 1.
\end{array}
\end{equation}
Regarding the constraint $C_{5a}$, it can be observed that its left-hand side is a non-convex function with respect to $\alpha_{k,n}$, which renders this constraint a non-convex constraint. Based on the characteristic of $C_{5a}$, by applying the first-order Taylor series expansion on $\alpha_{k,n}^2$, the lower bound of $\alpha_{k,n}^2$ can be derived as $\alpha_{k,n}^2\geq\bar\alpha_{k,n}^2 + 2\bar\alpha_{k,n}(\alpha_{k,n}-\bar\alpha_{k,n})$, the feasible set obtained always lies within the feasible set of $C_{5a}$.  Then, $C_{5a}$ can be rewritten as
\begin{equation}\label{eq21}
\begin{array}{l}
\bar C_{5a}:\alpha_{k,n}-(\bar\alpha_{k,n}^2 + 2\bar\alpha_{k,n}(\alpha_{k,n}-\bar\alpha_{k,n}))\leq 0,
\end{array}
\end{equation}
where $\bar\alpha_{k,n}$ is the previous iteration of $\alpha_{k,n}$.
\subsubsection{The Transformations of $C_2$ and $E_{\rm total}$}
It is worth noting that matrices $\boldsymbol{\rm \Lambda}_k$ and $\boldsymbol{\rm \Theta}_k$ in $C_2$ and $C_4$ are always coupled in the product form. Hence, we rewrite the product term $\boldsymbol{\rm \Lambda}_k\boldsymbol{\rm \Theta}_k$ as $\boldsymbol{\rm\bar \Theta}_k=\textrm{diag}(\boldsymbol{{\bar\theta}}_k)=\textrm{diag}(\rho_{k,1}^{\alpha_{k,1}}e^{j\theta_{k,1}},\cdots,\rho_{k,N}^{\alpha_{k,N}}e^{j\theta_{k,N}})\in\mathbb{C}^{N\times N}$, where $\boldsymbol{\rm\bar \theta}_k=[\bar\theta_{k,1},..,\bar\theta_{k,N}]^T\in \mathbb{C}^{N\times 1}$. However, $\boldsymbol{\rm A}_k\boldsymbol{\rm\bar \Theta}_k$ is still coupled. To decouple the coupled variables $\boldsymbol{\rm A}_k\boldsymbol{\rm\bar \Theta}_k$, we introduce a slack optimization variable $\boldsymbol{\rm u}_k=[u_{k,1},\cdots,u_{k,n},\cdots,u_{k,N}]^T\in \mathbb{C}^{N\times 1}$ and $\textrm{diag}(\boldsymbol{\rm u}_k^H)=\boldsymbol{\rm A}_k\boldsymbol{\rm\bar \Theta}_k$. Furthermore, we introduce a slack variable $\boldsymbol{\rm U}_k$, where
\begin{equation}\label{eq23}
\begin{array}{l}
	\boldsymbol{\rm U}_k=\boldsymbol{\rm u}_k\boldsymbol{\rm u}_k^H.
\end{array}
\end{equation}
The equation in (\ref{eq23}) can be equivalently rewritten as the following constraints\cite{h3}
\begin{equation}\label{eq25}
 C_{10a}:\left[                 
\begin{array}{cc}   
\boldsymbol{\rm U}_k& \boldsymbol{\rm u}_k\\  
\boldsymbol{\rm u}_k^H & 1\\  
\end{array}
\right] \succeq \boldsymbol{0},~ C_{10b}: {\rm Tr}(\boldsymbol{\rm U}_k-\boldsymbol{\rm u}_k\boldsymbol{\rm u}_k^H)\leq 0.      
\end{equation}
Based on the first-order Taylor expansion, the lower bound of ${\rm Tr}(\boldsymbol{\rm u}_k\boldsymbol{\rm u}_k^H)$ can be derived as
\begin{equation}\label{eq27}
\begin{array}{l}
{\rm Tr}(\boldsymbol{\rm u}_k\boldsymbol{\rm u}_k^H)\geq -\|\boldsymbol{\rm\bar u}_k\|^2+2{\rm Tr}(\boldsymbol{\rm\bar u}_k^H\boldsymbol{\rm u}_k),
\end{array}
\end{equation}
where $\boldsymbol{\rm\bar u}_k$ is the previous iteration of $\boldsymbol{\rm u}_k$. Thus, by substituting the lower bound in (\ref{eq27}) into $C_{10b}$, $C_{10b}$ can be rewritten as
\begin{equation}\label{eq29}
\begin{array}{l}
\bar C_{10b}:{\rm Tr}(\boldsymbol{\rm U}_k)\leq -\|\boldsymbol{\rm\bar u}_k\|^2+2{\rm Tr}(\boldsymbol{\rm\bar u}_k^H\boldsymbol{\rm u}_k).
\end{array}
\end{equation}
Then, $C_2$ and $E_k^{\textrm{total}}$ can be respectively rewritten as
\begin{equation}\label{eq32}
\begin{array}{l}
\!\!\!\!\!\!\bar C_2: p_k\|\boldsymbol{\rm A}_k\boldsymbol{\rm \Lambda}_k\boldsymbol{\rm \Theta}_k\boldsymbol{\rm h}_{{\rm r}, k}\|^2{+}\sigma^2\|\boldsymbol{\rm A}_k\boldsymbol{\rm \Lambda}_k\boldsymbol{\rm \Theta}_k\|^2{\leq} P_{\rm R}^{\max}\\
\!\!\!\!\!\!{\Rightarrow}  p_k\textrm{Tr}(\textrm{diag}(\boldsymbol{\rm h}_{{\rm r}, k})\boldsymbol{\rm U}_k \textrm{diag}(\boldsymbol{\rm h}_{{\rm r}, k})^H){+}\sigma^2\textrm{Tr}(\boldsymbol{\rm U}_k)   {\leq} P_{\rm R}^{\max},
\end{array}
\end{equation}
\begin{equation}\label{eq33}
\begin{array}{l}
\bar E^{\rm total}_k=t_kp_k+E_k^{\textrm{local}}+t_k[\sum\limits_{n=1}^N\alpha_{k,n}(P_{\textrm{C}}+P_{\textrm{DC}})\\
+(p_k\textrm{Tr}(\textrm{diag}(\boldsymbol{\rm h}_{{\rm r}, k})\boldsymbol{\rm U}_k \textrm{diag}(\boldsymbol{\rm h}_{{\rm r}, k})^H)+\sigma^2\textrm{Tr}(\boldsymbol{\rm U}_k))]+t_kP_k^{\rm pas}.
\end{array}
\end{equation}
\subsubsection{The Transformation of $C_4$} First, we introduce a slack variable $\boldsymbol{\rm O}_k$, where
\begin{equation}\label{eq24}
\begin{array}{l}
\boldsymbol{\rm O}_k=\boldsymbol{\rm o}_k\boldsymbol{\rm o}_k^H,
\end{array}
\end{equation}
where $\boldsymbol{\rm o}_k=[\boldsymbol{{\bar\theta}}_k;1]$.  $\boldsymbol{{\bar\Theta}}_k$ can be denoted by $\textrm{diag}([\boldsymbol{\rm O}_k]_{1:N,{N+1}})$, where $[\boldsymbol{\rm O}_k]_{1:N,{N+1}}=[[\boldsymbol{\rm O}_k]_{1,{N+1}};\cdots;[\boldsymbol{\rm O}_k]_{N,{N+1}}]$. The equations in (\ref{eq24}) can be equivalently rewritten as
\begin{equation}\label{eq26}
C_{10c}:\left[                 
\begin{array}{cc}   
\boldsymbol{\rm O}_k& \boldsymbol{\rm o}_k\\  
\boldsymbol{\rm o}_k^H & 1\\  
\end{array}
\right] \succeq \boldsymbol{0},~ C_{10d}: {\rm Tr}(\boldsymbol{\rm O}_k-\boldsymbol{\rm o}_k\boldsymbol{\rm o}_k^H)\leq 0.    
\end{equation}
Based on the first-order Taylor expansion, the lower bound of ${\rm Tr}(\boldsymbol{\rm o}_k\boldsymbol{\rm o}_k^H)$ can be derived as
\begin{equation}\label{eq28}
\begin{array}{l}
{\rm Tr}(\boldsymbol{\rm o}_k\boldsymbol{\rm o}_k^H)\geq -\|\boldsymbol{\rm\bar o}_k\|^2+2{\rm Tr}(\boldsymbol{\rm\bar o}_k^H\boldsymbol{\rm o}_k),
\end{array}
\end{equation}
where $\boldsymbol{\rm\bar o}_k$ are the previous iteration of $\boldsymbol{\rm o}_k$. Thus, by substituting the lower
bound in (\ref{eq28}) into $C_{10d}$, $C_{10d}$ can be rewritten as
\begin{equation}\label{eq30}
\begin{array}{l}
\bar C_{10d}:{\rm Tr}(\boldsymbol{\rm O}_k)\leq -\|\boldsymbol{\rm\bar o}_k\|^2+2{\rm Tr}(\boldsymbol{\rm\bar o}_k^H\boldsymbol{\rm o}_k).
\end{array}
\end{equation}
Then, the SINR expression can be rewritten as
\begin{equation}\label{eq31}
\begin{array}{l}
\bar\gamma_k=\frac{p_k|\boldsymbol{\rm w}_k^H(\boldsymbol{\rm h}_{{\rm d},k}+\boldsymbol{\rm H}^H\boldsymbol{\rm\bar \Theta}_k\boldsymbol{\rm h}_{{\rm r}, k})|^2}{\sigma^2|\boldsymbol{\rm w}_k^H\boldsymbol{\rm H}^H\textrm{diag}(\boldsymbol{\rm u}_k^H)|^2+\delta^2}\\
~~~=\frac{p_k\textrm{Tr}(\boldsymbol{\rm W}_k\boldsymbol{\rm H}_k\boldsymbol{\rm O}_k\boldsymbol{\rm H}_k^H)}{\sigma^2\textrm{Tr}(\textrm{diag}(\boldsymbol{\rm u}_k)\boldsymbol{\rm H}\boldsymbol{\rm W}_k\boldsymbol{\rm H}^H \textrm{diag}(\boldsymbol{\rm u}_k^H))  +\delta^2},
\end{array}
\end{equation}
where $\boldsymbol{\rm W}_k=\boldsymbol{\rm w}_k\boldsymbol{\rm w}_k^H$, $\boldsymbol{\rm H}_k=[\boldsymbol{\rm H}^H\textrm{diag}(\boldsymbol{\rm h}_{{\rm r}, k}),\boldsymbol{\rm h}_{{\rm d}, k}]$. Now, it can be observed that the numerator of the SINR expression is linear with respect to $\boldsymbol{ \rm O}_k$. However, the current form of the denominator is not tractable. Therefore, we apply the SVD method \cite{h4} to deal with the coupled variables, which facilitates the design of an efficient algorithm for the hybrid RIS, i.e.,  $\boldsymbol{\rm H}\boldsymbol{\rm W}_k\boldsymbol{\rm H}^H=\sum_{n=1}^N\chi_{k,n}\boldsymbol{\rm p}_{k,n}\boldsymbol{\rm q}_{k,n}^H$, where $\chi_{k,n}$, $\boldsymbol{\rm p}_{k,n}$, and $\boldsymbol{\rm q}_{k,n}$ are the singular values, the corresponding left and right singular vectors, respectively. Accordingly, we have the following equation
\begin{equation}\label{eq34}
	\begin{array}{l}
	~~~{\rm Tr}(\textrm{diag}(\boldsymbol{\rm u}_k)\boldsymbol{\rm H}\boldsymbol{\rm W}_k\boldsymbol{\rm H}^H \textrm{diag}(\boldsymbol{\rm u}_k^H)) \\
	={\rm Tr}(\sum\limits_{n=1}^N\chi_{k,n}\textrm{diag}(\boldsymbol{\rm p}_{k,n})\boldsymbol{\rm U}_k \textrm{diag}(\boldsymbol{\rm q}_{k,n}^H)).
	\end{array}
\end{equation}
Then, the SINR can be rewritten as
\begin{equation}\label{eq35}
\begin{array}{l}
\tilde\gamma_k=\frac{p_k\text{Tr}(\boldsymbol{\rm W}_k\boldsymbol{\rm H}_k\boldsymbol{\rm O}_k\boldsymbol{\rm H}_k^H)}{\sigma^2{\rm Tr}(\sum\limits_{n=1}^N\chi_{k,n}\textrm{diag}(\boldsymbol{\rm p}_{k,n})\boldsymbol{\rm U}_k \textrm{diag}(\boldsymbol{\rm q}_{k,n}^H))  {+}\delta^2}.\\
\end{array}
\end{equation}
Regarding the SINR expression, it can be observed that the optimization variables in the numerator and denominator of the SINR expression are coupled with each other. This motivates us to employ the SCA technique, which is a conservative and restrictive approximation, in a sense, the obtained solution is a feasible solution to problem (\ref{eq19}). Moreover, the SCA technique can guarantee to provide a stationary point, and the approximation accuracy depends on the initial point. Specifically, by using auxiliary variables to relax the SINR expression, the coupling between optimization variables can be decoupled well, i.e.,
\begin{equation}\label{eq36}
\begin{array}{l}
{C_{4a}}:p_k\text{Tr}(\boldsymbol{\rm W}_k\boldsymbol{\rm H}_k\boldsymbol{\rm O}_k\boldsymbol{\rm H}_k^H)\geq e^{u_k},
\end{array}
\end{equation}
\begin{equation}\label{eq37}
\begin{array}{l}
\!\!\!\!\!\!\!\!\!{C_{4b}}:\sigma^2{\rm Tr}(\sum\limits_{n=1}^N\chi_{k,n}\textrm{diag}(\boldsymbol{\rm p}_{k,n})\boldsymbol{\rm U}_k \textrm{diag}(\boldsymbol{\rm q}_{k,n}^H))  {+}\delta^2{\leq} e^{v_k}.
\end{array}
\end{equation}
Then, following the principle of the SCA method, we further approximate the non-convex constraint (\ref{eq37}) by its first-order Taylor expansion
\begin{equation}\label{eq38}
\begin{array}{l}
{\bar C_{4b}}:~\sigma^2{\rm Tr}(\sum\limits_{n=1}^N\chi_{k,n}\textrm{diag}(\boldsymbol{\rm p}_{k,n})\boldsymbol{\rm U}_k \textrm{diag}(\boldsymbol{\rm q}_{k,n}^H))  +\delta^2\\
~~~~~~~\leq e^{\bar v_k}({v_k}{-}{\bar v_k}{+}1),
\end{array}
\end{equation}
where $\bar v_k$ is the previous iteration of $v_k$. 
To make $C_8$ tractable, we adopt the same method as \cite{h5} to relax this constraint as
\begin{equation}\label{eq38-1}
\begin{array}{l}
\!\!\!\!\!\bar C_8:|[\textrm{diag}([\boldsymbol{\rm O}_k]_{1:N,{N+1}}){-}\textrm{diag}(\boldsymbol{\rm u}_k^H)]_{n,n}|{\leq}[\boldsymbol{\rm I}{-}\boldsymbol{\rm A}_k]_{n,n}.
\end{array}
\end{equation}
However, there still exists the non-convex equation constraint, and we adopt the big-M formulation \cite{h3} to decompose it. According to the additional convex constraints imposed, we can guarantee that the original equation holds regardless of the value of the discrete variable. Thus, $\textrm{diag}(\boldsymbol{\rm u}_k^H)=\boldsymbol{\rm A}_k\boldsymbol{\rm\bar \Theta}_k$ can be converted into a set of equivalent constraints as follows:
\begin{equation}\label{eq22}
\begin{array}{l}
\!\!\!\!C_{9a}:{\rm diag}([\boldsymbol{ \rm o}_k]_{1:N})-(\boldsymbol{ \rm I}_N-\boldsymbol{ \rm A}_k)\rho^{\max}\preceq {\rm diag}(\boldsymbol{ \rm u}_k^H),\\
\!\!\!\!C_{9b}:{\rm diag}(\boldsymbol{ \rm u}_k^H)\preceq {\rm diag}([\boldsymbol{ \rm o}_k]_{1:N})+(\boldsymbol{ \rm I}_N-\boldsymbol{ \rm A}_k)\rho^{\max},\\
\!\!\!\!C_{9c}:{\rm diag}(\boldsymbol{ \rm u}_k^H)\succeq \boldsymbol{ 0},~~C_{9d}:{\rm diag}(\boldsymbol{ \rm u}_k^H)\preceq \boldsymbol{ \rm A}_k\rho^{\max}\boldsymbol{ \rm I}_N,
\end{array}
\end{equation}
where $\rho^{\max}$ is an arbitrarily large constant. $C_{9a}$ and $C_{9b}$ are imposed to guarantee that $\textrm{diag}(\boldsymbol{\rm u}_k^H)=\boldsymbol{\rm A}_k\boldsymbol{\rm\bar \Theta}_k$ holds after reformulation.
Then, problem (\ref{eq19}) can be reformulated as
\begin{equation}\label{eq39}
\begin{array}{l}
\min\limits_{\mbox{\scriptsize$\begin{array}{c} 
		\boldsymbol{\rm O}_k,\boldsymbol{\rm o}_k,\boldsymbol{\rm U}_k,\boldsymbol{\rm u}_k\\
		u_k,v_k,\alpha_{k,n},f_k^{\textrm{local}},
		\end{array}$}} 
\!\!\!\sum\limits_{k=1}^K w_kt_k{+}\sum\limits_{k=1}^K(1{-}w_k)\psi \bar E^{\rm total}_k{\triangleq}\mathcal{F}\\
\!\!\!\!s.t.~{C_1},{\bar C_2},{C_{4a}},{\bar C_{4b}},\bar C_{5a},C_{5b},C_6,{C_7},{\bar C_8},C_{9a}{-}C_{9d},\\
~~~C_{10a},\bar C_{10b},C_{10c},\bar C_{10d},{\bar C_4}:Bt_k\frac{u_k-v_k}{\ln2}\geq \beta_kS_k.
\end{array}
\end{equation}
Problem (\ref{eq39}) is a convex optimization problem and can be solved directly by the standard convex optimization techniques. The proposed algorithm based on mode switching is shown in \textbf{Algorithm 1}.
\begin{spacing}{1.00}
	\floatname{algorithm}{Algorithm}
	\renewcommand{\algorithmicrequire}{\textbf{Input:}}
	\renewcommand{\algorithmicensure}{\textbf{Output:}}
	\begin{algorithm}[!t]
		\small
		\caption{: The Joint Optimization Algorithm Based on Mode Switching}
		\begin{algorithmic}[1]
			\State Initialize system parameters: $K$, $M$, $N$, $P_{\rm R}^{\max}$, $E_k^{\max}$, $S_k$, $C_k$, $f_k^{\max}$, $\delta^2$, $\sigma^2$, $P_{\rm C}$, $P_{\rm DC}$, $\psi$, $\kappa$, $B$, $\bar \alpha_{k,n}$, $\boldsymbol{{\rm\bar u}}_k$, $\boldsymbol{{\rm\bar o}}_k$, $\bar v_k$;
			\State Set the maximum iteration number $L_{\max}$ and the convergence accuracy $\epsilon$, set the initial iteration index $l=0$;
			\While{$l\leq L_{\max}$}
			\State Initialize $p_k$, $\boldsymbol{\rm A}_k$, $\boldsymbol{\rm \Lambda}_k$, and $\boldsymbol{\rm \Theta}_k$, calculate $\boldsymbol{{\rm w}}_{k}$ via (\ref{eq13});
			\State Given $\boldsymbol{\rm w}_k$, $\boldsymbol{\rm A}_k$, $\boldsymbol{\rm \Lambda}_k$, $\boldsymbol{\rm \Theta}_k$, and $f_k^{\rm local}$, calculate $p_k$, $t_k$, and $\beta_k$ via (\ref{eq18});
			\Repeat
			\State Given $p_k$, $t_k$, and $\beta_k$, calculate $\boldsymbol{\rm \Theta}_k$, $\alpha_{k,n}$, $\boldsymbol{\rm \Lambda}_k$, and $f_k^{\textrm{local}}$ via (\ref{eq39});
			\Until{Convergence};
			\If{$|\mathcal{F}(l)-\mathcal{F}(l-1)|\leq\epsilon$, where $\mathcal{F}(l)=\sum_{k=1}^K w_kt_k(l)+\sum_{k=1}^K(1-w_k)\psi E^{\rm total}_k(l)$}
			\State \textbf{break}.
			\Else
			\State $l=l+1$;
			\EndIf
			\EndWhile		
		\end{algorithmic}
	\end{algorithm}
\end{spacing} \vspace{-10pt}
\subsection{Convergence Analysis}
The convergence of the proposed algorithm can be guaranteed as follows: Problem (\ref{eq18}) and problem (\ref{eq39}) are standard convex optimization problems, thus the solutions obtained in each iteration are all optimal solutions to the respective sub-problems. Let $t_k^{(l)}$, $p_k^{(l)}$, $\boldsymbol{\rm \Theta}_k^{(l)}$, $\beta_k^{(l)}$,
$\boldsymbol{\rm w}_k^{(l)}$,
$\alpha_{k,n}^{(l)}$, $\boldsymbol{\rm \Lambda}_k^{(l)}$, $(f_k^{\textrm{local}})^{(l)}$ be the solutions obtained in the $l$-th iteration, defining the objective function as $\mathcal{F}(\boldsymbol{\rm w}_k,t_k,p_k,\beta_k, \boldsymbol{\rm \Theta}_k,
\alpha_{k,n},\boldsymbol{\rm \Lambda}_k,f_k^{\textrm{local}})\triangleq\sum_{k=1}^K w_kt_k+\sum_{k=1}^K(1-w_k)\psi E^{\rm total}_k$. Then, we have
\begin{equation}\label{eq40}
\begin{array}{l}
\mathcal{F}(\boldsymbol{\rm w}_k^{(l)},t_k^{(l)},p_k^{(l)},\beta_k^{(l)},
\boldsymbol{\rm \Theta}_k^{(l)},\alpha_{k,n}^{(l)},\boldsymbol{\rm \Lambda}_k^{(l)},(f_k^{\textrm{local}})^{(l)})\\
{\geq} \mathcal{F}(\boldsymbol{\rm w}_k^{(l+1)},t_k^{(l)},p_k^{(l)},\beta_k^{(l)},
\boldsymbol{\rm \Theta}_k^{(l)},\alpha_{k,n}^{(l)},\boldsymbol{\rm \Lambda}_k^{(l)},(f_k^{\textrm{local}})^{(l)})\\
{\geq} \mathcal{F}(\boldsymbol{\rm w}_k^{(l+1)},t_k^{(l+1)},p_k^{(l+1)},\beta_k^{(l+1)},
\boldsymbol{\rm \Theta}_k^{(l)},\alpha_{k,n}^{(l)},\boldsymbol{\rm \Lambda}_k^{(l)},(f_k^{\textrm{local}})^{(l)})\\
{\geq} \mathcal{F}(\boldsymbol{\rm w}_k^{(l{+}1)}\!\!,\!t_k^{(l{+}1)}\!\!,\!p_k^{(l{+}1)}\!\!,\!\beta_k^{(l{+}1)}\!\!,\!\boldsymbol{\rm \Theta}_k^{(l{+}1)}\!\!,\!\alpha_{k,n}^{(l{+}1)}\!\!,\!\boldsymbol{\rm \Lambda}_k^{(l{+}1)}\!\!,\!\!(f_k^{\textrm{local}})^{(l{+}1)}).\\
\end{array}
\end{equation}
Based on the step 4 in Algorithm 1, $\boldsymbol{\rm w}_k^{(l+1)}$ can be obtained for given other variables, we have $\mathcal{F}(\boldsymbol{\rm w}_k^{(l)},t_k^{(l)},p_k^{(l)},\beta_k^{(l)},
\boldsymbol{\rm \Theta}_k^{(l)},\alpha_{k,n}^{(l)},\boldsymbol{\rm \Lambda}_k^{(l)},(f_k^{\textrm{local}})^{(l)})
{\geq} \mathcal{F}(\boldsymbol{\rm w}_k^{(l+1)},\\
t_k^{(l)},p_k^{(l)},\beta_k^{(l)},
\boldsymbol{\rm \Theta}_k^{(l)},\alpha_{k,n}^{(l)},\boldsymbol{\rm \Lambda}_k^{(l)},(f_k^{\textrm{local}})^{(l)})$. Similarly, based on steps 5 and 7 in Algorithm 1, we can obtain equation (\ref{eq40}). The obtained optimal objective value of Algorithm 1 at each iteration is non-increasing or remains unchanged. Besides, the objective value is non-negative and bounded. Thus, Algorithm 1 converges to a locally optimal point after several iterations.

\subsection{Computational Complexity Analysis}
The joint optimization algorithm based on mode switching requires solving problem (\ref{eq18}) and problem (\ref{eq39}). For problem (\ref{eq18}), there exist $6K$ linear matrix inequality (LMI) constraints of size $1$. Let $\epsilon_1$ be the iteration accuracy of problem (\ref{eq18}), the number of iterations is on the order of $\ln(\frac{1}{\epsilon_1})\sqrt{6K}$. There exist $3K$ real variables, the number of optimization variables of problem (\ref{eq18}) is on the order of $3K$. Thus, the overall computational complexity of problem (\ref{eq18}) is calculated as $o_1=\mathcal{O}\{ \ln(\frac{1}{\epsilon_1})\sqrt{6K}(9K^2(2+9K))     \}$.
For problem (40), there exist $K$ LMI constraints of size $N+2$, $K$ LMI constraints of size $N+1$, and $10K+3KN$ LMI constraints of size $1$. Let $\epsilon_2$ be the iteration accuracy of problem (40), the number of iterations is on the order of $a_1=\ln(\frac{1}{\epsilon_2})\sqrt{K(9N+13)}$. There exist $3K+KN$ real variables, $2K$ matrix variables of size $N+1$, and $2K$ matrix variables of size $N$. The number of optimization variables of problem (40) is on the order of $a_2=2K(N+1)^2+2KN^2+K(N+3)$. Thus, the overall computational complexity of problem (40) is calculated as $o_2=\mathcal{O}\{a_1(a_2b_1+a_2^2b_2+a_2^3)\}$, where $b_1=K(N+1)^3+K(N+2)^3+K(7N+9)$ and $b_2=K(N+1)^2+K(N+2)^2+K(7N+9)$.
Thus, the computational complexity of Algorithm 1 is $o_1+o_2$.
\section{The Tradeoff Between the Number of Active and Passive Reflecting Units}
In the previous section, we initially formulate a joint optimization problem that aims to achieve a tradeoff between latency and energy consumption by optimizing multiple system parameters. We apply an AO algorithm and some SCA methods to obtain the sub-optimal solutions numerically, which, however, may not gain some unique insights to guide actual deployment. We may not clearly observe how Algorithm 1 works and directly analyze the relationship between the number of active and passive units due to the lack of closed-form solutions to problem (\ref{eq12}). In order to gain more insight into the problem and provide a more intuitive analysis of the mode-switching mechanism, we consider two special cases for problem (\ref{eq12}) in this section, namely latency minimization and energy consumption minimization, which are also important optimization problems in the RIS-aided MEC systems and can effectively reflect the characteristics of the mode-switching mechanism. This leads to an algorithm in which each iteration is performed in closed form rather than solving a convex optimization problem numerically, which is often more desirable than the numerical solution method introduced in Section III.  To facilitate our analysis, we assume that the amplification factor of each reflecting unit is the same \cite{e2}, each user performs the task offloading at a maximum power $P_k^{\max}$, and the AP is equipped with a single antenna, i.e., $\boldsymbol{\rm H}\in\mathbb{C}^{N\times M}\rightarrow\boldsymbol{\rm h}\in\mathbb{C}^{N\times 1}$. Moreover, we ignore the direct link and assume that all task is offloaded to the MEC due to the reason that the direct link and local computing is independent of the number of reflecting units. Meanwhile, we assume that each channel involved is a line-of-sight channel with a large-scale path loss. In what follows, we will analyze the latency minimization problem and energy consumption minimization problem, respectively.

\subsection{Latency Minimization}
In this subsection, we focus on the tradeoff relationship between active and passive reflecting units in the latency minimization problem. Thus, the original optimization problem can be degraded to the following latency minimization problem when the tradeoff factor $w_k=1$ holds
\begin{equation}\label{eq41}
\begin{array}{l}
\min\limits_{\mbox{\scriptsize$\begin{array}{c} 
		\boldsymbol{\rm \Theta}_k,t_k\\
		\alpha_{k,n},\boldsymbol{\rm \Lambda}_k
		\end{array}$}} 
\sum\limits_{k=1}^K t_k\\
s.t.~
{C_2}:~ P_k^{\max}\|\boldsymbol{\rm A}_k\boldsymbol{\rm \Lambda}_k\boldsymbol{\rm \Theta}_k\boldsymbol{\rm h}_{{\rm r}, k}\|^2+\sigma^2\|\boldsymbol{\rm A}_k\boldsymbol{\rm \Lambda}_k\boldsymbol{\rm \Theta}_k\|^2\leq P_{\rm R}^{\max},\\
~~~~~{C_4}:~t_kB\log_2(1+\gamma_k)\geq S_k,\\
~~~~~{C_5}:~\alpha_{k,n}\in\{0,1\},\\
~~~~~C_6:~t_k\leq  T_k^{\max},
\end{array}
\end{equation}
where
\begin{equation}\label{eq42}
\begin{array}{l}
\gamma_k=\frac{P_k^{\max}|\boldsymbol{\rm h}^H\boldsymbol{\rm \Lambda}_k\boldsymbol{\rm \Theta}_k\boldsymbol{\rm h}_{{\rm r}, k}|^2}{\sigma^2\|\boldsymbol{\rm h}^H\boldsymbol{\rm A}_k\boldsymbol{\rm \Lambda}_k\boldsymbol{\rm \Theta}_k\|^2+\delta^2}.
\end{array}
\end{equation}
In the following, we first drop $C_6$ to make problem (\ref{eq41}) more treatable and then determine its feasible condition at the end. Thus, we have
\begin{equation}\label{eq43}
\begin{array}{l}
\min\limits_{\mbox{\scriptsize$\begin{array}{c} 
		\boldsymbol{\rm \Theta}_k,t_k\\
		\alpha_{k,n},\boldsymbol{\rm \Lambda}_k
		\end{array}$}} 
\sum\limits_{k=1}^K t_k\\
s.t.~{C_2},{C_4},{C_5}.
\end{array}
\end{equation}
There is no co-channel interference among users since we consider a user offloading strategy based on a TDMA manner. Besides, it can be observed from (\ref{eq43}) that each constraint is individual for each user. Thus, there is no resource competition among users. To facilitate algorithm design, we can decompose problem (\ref{eq43}) by solving the following problem 
\begin{equation}\label{eq44}
\begin{array}{l}
\min\limits_{\mbox{\scriptsize$\begin{array}{c} 
		\boldsymbol{\rm \Theta}_k,t_k\\
		\alpha_{k,n},\boldsymbol{\rm \Lambda}_k
		\end{array}$}} 
t_k\\
s.t.~{C_2},{C_4},{C_5}.
\end{array}
\end{equation}
Given an input data size $S_k$, the minimum latency can be obtained when the equation holds in $C_4$. Thus, problem (\ref{eq44}) can be equivalently transformed into
\begin{equation}\label{eq45}
\begin{array}{l}
\max\limits_{\mbox{\scriptsize$\begin{array}{c} 
		\boldsymbol{\rm \Theta}_k,
		\alpha_{k,n},\boldsymbol{\rm \Lambda}_k
		\end{array}$}} 
\gamma_k\\
s.t.~{C_2},{C_5}.\\
\end{array}
\end{equation}
Problem (\ref{eq45}) is challenging to solve since the phase-shift matrix, the mode-switching matrix, the transmission time, and the amplification factor are coupled in the objective function and constraints. Thus, we first obtain the closed-form solution of the phase shift. 
\subsubsection{Phase-Shift Optimization}
Given the mode-switching factor, the amplification factor, and the transmission time, it can be easily shown that the SINR in (\ref{eq45}) is maximized when the optimal phase-shift matrix of the RIS aligns the cascaded user-RIS-AP channel\cite{i2,i2-1}, i.e.,
\begin{equation}\label{eq46}
\begin{array}{l}
\theta_{k,n}^*={\rm arg}([\boldsymbol{\rm h}]_n)-{\rm arg}([\boldsymbol{\rm h}_{{\rm r},k}]_n),
\end{array}
\end{equation}
where ${\rm arg}(\cdot)$ denotes the angle of a complex number. Then, by substituting the optimal phase shift in (\ref{eq46}) into the SINR expression and $C_2$, we have
\begin{equation}\label{eq47}
\begin{array}{l}
\bar\gamma_k=\frac{P_k^{\max}|\boldsymbol{\rm h}^H\boldsymbol{\rm \Lambda}_k\boldsymbol{\rm \Theta}_k\boldsymbol{\rm h}_{{\rm r}, k}|^2}{\sigma^2\|\boldsymbol{\rm h}^H\boldsymbol{\rm A}_k\boldsymbol{\rm \Lambda}_k\boldsymbol{\rm \Theta}_k\|^2{+}\delta^2}=\frac{P_k^{\max}|\boldsymbol{\rm h}^H\boldsymbol{\rm \Lambda}_k\boldsymbol{\rm \Theta}_k\boldsymbol{\rm h}_{{\rm r}, k}|^2}{\sigma^2\sum\limits_{n=1}^N(\alpha_{k,n}\rho_{k,n}^{\alpha_{k,n}}|h_n|)^2{+}\delta^2}\\
\overset{(a)}{=}\frac{P_k^{\max}(\sum\limits_{n=1}^N\rho_{k,n}^{\alpha_{k,n}}|h_n||h_{{\rm r},k}^n|)^2}{\sigma^2\sum\limits_{n=1}^N(\alpha_{k,n}\rho_{k,n}^{\alpha_{k,n}}|h_n|)^2+\delta^2}\overset{(b)}{\geq}\frac{P_k^{\max}|f_k|^2(\sum\limits_{n=1}^N\rho_{k,n}^{\alpha_{k,n}})^2}{\sigma^2|h|^2\sum\limits_{n=1}^N(\alpha_{k,n}\rho_{k,n}^{\alpha_{k,n}})^2+\delta^2},
\end{array}
\end{equation}
\begin{equation}\label{eq48}
\begin{array}{l}
\!\!\!\!\bar C_2:P_k^{\max}|{h}_{{\rm r}, k}|^2\sum\limits_{n=1}^N(\alpha_{k,n}\rho_{k,n}^{\alpha_{k,n}})^2{+}\sigma^2\sum\limits_{n=1}^N(\alpha_{k,n}\rho_{k,n}^{\alpha_{k,n}})^2{\leq} P_{\rm R}^{\max},\\
\end{array}
\end{equation}
where $(a)$ utilizes the optimal design of $\boldsymbol{{\Theta}}_k$, $(b)$ utilizes $|f_k|=\min\{|h_n||h_{{\rm r},k}^n|\}$, $|h|=\max\{|h_n|\}$, and $|h_{{\rm r},k}|=\max\{|{{ h}}_{{\rm r},k}^n|\}$. $h_n$ and ${{ h}}_{{\rm r},k}^n$ are the $n$-th elements of $\boldsymbol{\rm h}$ and $\boldsymbol{\rm h}_{{\rm r},k}$, respectively. Furthermore, we reconstruct the SINR expression as follows
\begin{equation}\label{eq48}
\begin{array}{l}
\tilde\gamma_k=\frac{P_k^{\max}|f_k|^2(\sum\limits_{n=1}^N\alpha_{k,n}\rho_{k,n}+N-\sum\limits_{n=1}^N\alpha_{k,n})^2}{\sigma^2|h|^2\sum\limits_{n=1}^N\alpha_{k,n}\rho_{k,n}^2+\delta^2},
\end{array}
\end{equation}
where $\sum_{n=1}^N\alpha_{k,n}\rho_{k,n}$ denotes active unit component and $N-\sum_{n=1}^N\alpha_{k,n}$ denotes the passive unit component.
\subsubsection{Amplification Factor Optimization}
Next, we optimize the amplification factor given the phase shift, the transmission time, and the number of active units. Problem (\ref{eq45}) can be reduced to
\begin{equation}\label{eq53}
\begin{array}{l}
\max\limits_{\mbox{\scriptsize$\begin{array}{c} 
		\rho_k
		\end{array}$}} 
~~\tilde\gamma_k\\
s.t.~{\tilde C_2}.
\end{array}
\end{equation}
Then, we have the following Proposition.

\textit{\textbf{Proposition 1:}} When $N_{\rm act}>0$, the amplification factor for each active unit at the $k$-th time slot is given by
\begin{equation}\label{eq54}
\begin{array}{l}
\rho_k^*=\sqrt\frac{P_{\rm R}^{\max}}{( P_k^{\max}|h_{{\rm r}, k}|^2+\sigma^2)N_{\rm act}}.
\end{array}
\end{equation}
\begin{proof}
	See Appendix A.
\end{proof}
\textbf{\textit{Remark 1:}} Proposition 1 shows that there exists a tradeoff between the amplification factor and the number of active units. The amplification factor is decreased by increasing transmit power $P_k^{\max}$ and the number of active units $N_{\rm act}$, and increases with the increasing power threshold of the RIS $P_{\rm R}^{\max}$.

\subsubsection{Number Optimization}
Given the transmission time, the optimal phase shift, and the amplification factor, we focus on optimizing the number of active units to maximize the offloading rate. Then, problem (\ref{eq45}) can be reduced to
\begin{equation}\label{eq55}
\begin{array}{l}
\max\limits_{\mbox{\scriptsize$\begin{array}{c} 
		N_{\rm act}
		\end{array}$}} 
\frac{P_k^{\max}|f_k|^2(N_{\rm act}\rho_{k}{+}N{-}N_{\rm act})^2}{\sigma^2|h|^2N_{\rm act}\rho_{k}^2{+}\delta^2}\\
s.t.~{\bar C_2}: P_k^{\max}|{h}_{{\rm r}, k}|^2N_{\rm act}\rho_{k}^2+\sigma^2N_{\rm act}\rho_{k}^2\leq P_{\rm R}^{\max},\\
~~~~~{\bar C_5}:0\leq N_{\rm act}\leq N.\\
\end{array}
\end{equation}
Then, we substitute $\rho_k^*$ in (\ref{eq54}) into problem (\ref{eq55}), and problem (\ref{eq55}) can be reduced to 
	\begin{equation}\label{eq56}
	\begin{array}{l}
	\max\limits_{\mbox{\scriptsize$\begin{array}{c} 
			N_{\rm act}
			\end{array}$}} 
	\frac{P_k^{\max}|f_k|^2(\sqrt {\frac{P_{\rm R}^{\max}N_{\rm act}}{ P_k^{\max}|h_{{\rm r},k}|^2+\sigma^2}}+N-N_{\rm act}  )^2    }{\sigma^2|h|^2\frac{P_{\rm R}^{\max}}{ P_k^{\max}|h_{{\rm r},k}|^2+\sigma^2}+\delta^2}\\
	s.t.~\bar C_{5}:0\leq N_{\rm act}\leq N.\\
	\end{array}
	\end{equation}
Then, we have the following Proposition.

\textit{\textbf{Proposition 2:}} When $P_{\rm R}^{\max}> 4N\sigma^2$, the closed-form solution of $N_{\rm act}$ is given by (\ref{eq57}).
\begin{figure*}
	\vspace{-20pt}
	\begin{equation}\label{eq57}
	\begin{array}{l}
	\left\{
	\begin{array}{l}
	N_{\rm act}^*=N,N_{\rm pas}^*=0,~~~~~~~~~~~~~~~~~~~~~~~~~~~~~~~~~~~~~~~~~~~~~~~~~~~~~~~~~|h_{{\rm r},k}|\leq\sqrt{\frac{P_{\rm R}^{\max}}{4N P_k^{\max}}-\frac{\sigma^2}{ P_k^{\max}}},\\
	N_{\rm act}^*=\frac{P_{\rm R}^{\max}}{4( P_k^{\max}|h_{{\rm r},k}|^2+\sigma^2)},N_{\rm pas}^*=N-\frac{P_{\rm R}^{\max}}{4( P_k^{\max}|h_{{\rm r},k}|^2+\sigma^2)},~~~~~~~~~~~~~~~ |h_{{\rm r},k}|>\sqrt{\frac{P_{\rm R}^{\max}}{4N P_k^{\max}}-\frac{\sigma^2}{ P_k^{\max}}}.\\
	\end{array}
	\right.
	\end{array}
	\end{equation}
	\hrulefill 
\end{figure*}
\begin{proof}
	See Appendix B.
\end{proof}

\textit{\textbf{Remark 2:}} Proposition 2 shows that the number of active and passive reflecting units is mainly related to the transmit power, the channel condition between the user and the RIS, and the power threshold of the RIS. All reflecting units should be switched to the active mode when the channel condition between users and the RIS is poor, i.e., $|h_{{\rm r},k}|\leq\sqrt{\frac{P_{\rm R}^{\max}}{4N P_k^{\max}}-\frac{\sigma^2}{ P_k^{\max}}}$, which maximally compensates for performance degradation caused by the poor channel condition. There exists a tradeoff between the number of active units and the number of passive units when the channel condition between users and the RIS is good enough, i.e., $|h_{{\rm r},k}|>\sqrt{\frac{P_{\rm R}^{\max}}{4N P_k^{\max}}-\frac{\sigma^2}{ P_k^{\max}}}$. This is because when the channel condition is large enough, the useful signal power will increase, but the noise power at the RIS is also increased by the active reflecting unit. Therefore, some reflecting units will be switched to passive mode to ensure optimal performance and optimal amplification performance under a limited power threshold.

\textit{\textbf{Remark 3:}} The exploration of number configuration for the RIS is also involved in \cite{i2-2} and \cite{i2-3}, where the number of reflecting units for the passive RIS is optimized to guarantee the quality of service requirements of users. On this basis, we further explore the number of active and passive units for the hybrid RIS. It is worth noting that our results in Section III-A are different from those in \cite{f2} in the sense that we emphasize the relationship between the number configuration and channel conditions while \cite{f2} depicts the relationship between the element allocation and the total power budget.
\subsubsection{Feasible Condition for Transmission Time} So far, we have obtained the closed-form solutions for the phase shift, the amplification factor, and the number of active units. In this subsection, we will discuss the feasible condition for transmission time. When the optimization variables are obtained, problem (\ref{eq41}) is feasible when the following condition holds, i.e.,
\begin{equation}\label{eq58}
\begin{array}{l}
T_k^{\max}{\geq}
t_k^*{=}\frac{S_k}{B\log_2\left(1{+}\frac{P_k^{\max}|f_k|^2(N_{\rm act}^*\rho_{k}^*+N-N_{\rm act}^*)^2}{\sigma^2|h|^2N_{\rm act}^*(\rho_{k}^*)^2+\delta^2}\right)}.
\end{array}
\end{equation}
\begin{proof}	
	$C_4$ is the lower-bound constraint of the offloading time, and $C_6$ is the upper-bound constraint of the offloading time. Based on the previous derivation, we have obtained closed-form solutions for phase shift $\theta_{k,n}^*$, amplification factor $\rho_k^*$, and number of active and passive reflecting units $N_{\rm act}^*$, $N_{\rm pas}^*$. Then, given an input data size $S_k$, the minimum latency can be obtained when the equation holds in $C_4$. Thus, based on $C_4$, substituting the obtained solutions into the SINR expression, we have $R_k=B\log_2\left(1{+}\frac{P_k^{\max}|f_k|^2(N_{\rm act}^*\rho_{k}^*+N-N_{\rm act}^*)^2}{\sigma^2|h|^2N_{\rm act}^*(\rho_{k}^*)^2+\delta^2}\right)$. Then,  we can also obtain the closed-form solution of the offloading time $t_k^*=\frac{S_k}{B\log_2\left(1{+}\frac{P_k^{\max}|f_k|^2(N_{\rm act}^*\rho_{k}^*+N-N_{\rm act}^*)^2}{\sigma^2|h|^2N_{\rm act}^*(\rho_{k}^*)^2+\delta^2}\right)}$. Next, we consider the upper-bound constraint of the offloading time.  When the following condition holds: $t_k^*\leq T_k^{\max}$, problem (\ref{eq41}) is feasible, which completes the proof. 
\end{proof}
The number tradeoff of active and passive units for latency minimization is summarized in \textbf{Algorithm 2}.
\begin{spacing}{1.00}
	\floatname{algorithm}{Algorithm}
	\renewcommand{\algorithmicrequire}{\textbf{Input:}}
	\renewcommand{\algorithmicensure}{\textbf{Output:}}
	\begin{algorithm}[!t]
		\small
		\caption{: The Number Tradeoff of Active and Passive Units for Latency Minimization}
		\begin{algorithmic}[1]
			\State Initialize system parameters: $K$, $N$, $P_{\rm R}^{\max}$, $P_k^{\max}$, $S_k$, $\delta^2$, $\sigma^2$, $P_{\rm C}$, $P_{\rm DC}$, $B$;
			\State Calculate the phase shift $\theta_{k,n}^*$ by (\ref{eq46});
			\State Substitute the phase shift $\theta_{k,n}^*$ into problem (\ref{eq45}), calculate the amplification factor $\rho_k^*$ by (\ref{eq54});
			\State Substitute the phase shift $\theta_{k,n}^*$ and the amplification factor $\rho_k^*$ into problem (\ref{eq45}), calculate the number of active and passive units $N_{\rm act}^*$ and $N_{\rm pas}^*$ by (\ref{eq57});
			\If{The transmission time satisfies condition (\ref{eq58})}
			\State Output $\theta_{k,n}^*$, $\rho_k^*$, $N_{\rm act}^*$, and $N_{\rm pas}^*$;
			\Else
			\State Problem infeasible.
			\EndIf	
		\end{algorithmic}
	\end{algorithm}
\end{spacing}

\vspace{-15pt}
\subsection{Energy Consumption Minimization}
Different from the latency minimization problem in the previous subsection, this subsection focuses on the discussion of the tradeoff between active and passive reflecting units under the objective of energy consumption minimization. In particular, when the tradeoff factor $w_k=0$ holds, the original optimization problem can be degraded to
\begin{equation}\label{eq59}
\begin{array}{l}
\min\limits_{\mbox{\scriptsize$\begin{array}{c} 
		\boldsymbol{\rm \Theta}_k,t_k\\
		\alpha_{k,n},\boldsymbol{\rm \Lambda}_k
		\end{array}$}} 
\sum\limits_{k=1}^K(t_kP_k^{\max}+t_kP_k^{\rm act}+t_kP_k^{\rm pas})\\
s.t.~{C_2},C_4,{C_5},C_6.
\end{array}
\end{equation}
Similarly, the optimization variables are highly coupled, such as $t_k$, $\boldsymbol{\rm \Theta}_k$, etc. It is challenging to directly find optimal solutions to this problem. Next, we will explore the closed-form solutions to the above non-convex problem.
\subsubsection{Phase-Shift Optimization}
Given the mode-switching factor, the amplification factor, and the transmission time, the optimal phase shift can be denoted as
\begin{equation}\label{eq60}
\begin{array}{l}
\theta_{k,n}^*={\rm arg}([\boldsymbol{\rm h}]_n)-{\rm arg}([\boldsymbol{\rm h}_{{\rm r},k}]_n),
\end{array}
\end{equation}
where the optimal phase shift of the RIS has a similar form as the latency minimization problem.
\begin{proof}
	See Appendix C.
\end{proof}

\subsubsection{The Transmission Time Optimization}
Given any feasible amplification factor, the mode-switching factor, and the phase shift, problem (\ref{eq59}) can be reduced to
\begin{equation}\label{eq61}
\begin{array}{l}
\min\limits_{\mbox{\scriptsize$\begin{array}{c} 
		t_k
		\end{array}$}} 
\sum\limits_{k=1}^K(t_k P_k^{\max}+t_kP_k^{\rm act}+t_kP_k^{\rm pas})\\
s.t.~C_4,C_6.
\end{array}
\end{equation}
It can be observed that energy consumption is an increasing function in $t_k$, the minimum energy consumption can be achieved at the lower bound of $t_k$, then we need to find the lower bound of $t_k$. According to $C_4$ and ($\ref{eq60}$), we can obtain the lower bound of $t_k$, i.e.,
\begin{equation}\label{eq62}
\begin{array}{l}
t_k^{\rm L}=\frac{S_k}{B\log_2(1{+}\frac{P_k^{\max}|f_k|^2(N_{\rm act}\rho_k+N-N_{\rm act})^2}{\sigma^2|h|^2N_{\rm act}\rho_k^2+\delta^2})}.
\end{array}
\end{equation}
Then, the closed-form solution of the transmission time can be expressed as
\begin{equation}\label{eq63}
t_k^*=\left\{
\begin{array}{l}
t_k^{\rm L}, ~~~~~~~~~~~~t_k^{\rm L}\leq T^{\max},\\
\textrm{Infeasible}, ~~~\textrm{Otherwise}.
\end{array}
\right.
\end{equation}

\subsubsection{Amplification Factor Optimization} Given the number of active units, the phase shift, and the transmission time, the energy consumption minimization can be rewritten as
\begin{equation}\label{eq64}
\begin{array}{l}
\min\limits_{\mbox{\scriptsize$\begin{array}{c} 
		\rho_{k}
		\end{array}$}} 
\sum\limits_{k=1}^Kt_k^*(P_k^{\max}+P_k^{\max}|{h}_{{\rm r}, k}|^2N_{\rm act}\rho_k^{2}{+}\sigma^2N_{\rm act}\rho_k^{2}\\
~~~~~+NP_{\rm C}{+}N_{\rm act}P_{\rm DC})\\
s.t.
{\tilde C_2}:~ P_k^{\max}|{h}_{{\rm r}, k}|^2N_{\rm act}\rho_{k}^2{+}\sigma^2N_{\rm act}\rho_{k}^2{\leq} P_{\rm R}^{\max},\\
~~~~\tilde C_4:~\frac{P_k^{\max}|f_k|^2(N_{\rm act}\rho_k+N-N_{\rm act})^2}{\sigma^2|h|^2N_{\rm act}\rho_k^2+\delta^2}\geq \hat S_k,
\end{array}
\end{equation}
where $\hat S_k=2^{\frac{S_k}{t_k^*B}}-1$. Note that energy consumption is an increasing function in $\rho_k$, we need to find the lower bound of $\rho_k$. Then, defining $\rho^{\min}\geq1$ as the minimum amplification factor, the lower bound of $\rho_k$ can be obtained in (\ref{eq65}) by analyzing $\tilde C_4$, where
\begin{figure*}\vspace{-20pt}
	\begin{equation}\label{eq65}
	\small
	\rho_k^{\rm L}=\left\{
	\begin{array}{l}
	\max\left\{\frac{-b_k{+}\sqrt{(b_k)^2{-}4\sigma^2\delta^2|h|^2N_{\rm act}\hat S_k^2}}{{-}2\sigma^2|h|^2N_{\rm act}\hat S_k},\rho^{\min}\right\},~~~~b_k^2{\geq}4\sigma^2\delta^2|h|^2N_{\rm act}\hat S_k^2~\&~ \rho^{\min}{\leq}\frac{{-}b_k{-}\sqrt{(b_k)^2{-}4\sigma^2\delta^2|h|^2N_{\rm act}\hat S_k^2}}{{-}2\sigma^2|h|^2N_{\rm act}\hat S_k},\\
	\textrm{Infeasible}, ~~~~~~~~~~~~~~~~~~~~~~~~~~~~~~~~~~~~~~~~~~~~~~~~~~~~\textrm{Otherwise}.
	\end{array}
	\right.
	\end{equation}
	\hrulefill
\end{figure*}
\begin{equation}\label{eq66}
\begin{array}{l}
b_k=4P_k^{\max}|f_k|^2N_{\rm act}(N-N_{\rm act}).
\end{array}
\end{equation} 
\begin{proof}
	See Appendix D.
\end{proof}

Besides, based on $\tilde C_2$, we can obtain the upper bound of $\rho_k$, which can be expressed as
\begin{equation}\label{eq67}
\begin{array}{l}
\rho_k^{\rm U}=\sqrt\frac{P_{\rm R}^{\max}}{( P_k^{\max}|h_{{\rm r}, k}|^2+\sigma^2)N_{\rm act}}.
\end{array}
\end{equation} 
By comparing (\ref{eq65}) and (\ref{eq67}) and we have the following Proposition.

\textit{\textbf{Proposition 3:}} The closed-form solution of the amplification factor can be obtained when the following condition holds.
\begin{equation}\label{eq68}
\rho_k^*=\left\{
\begin{array}{l}
\rho_k^{\rm L},~~~~~~~~~~\rho_k^{\rm L}\leq\rho_k^{\rm U},\\
\textrm{Infeasible},~~\textrm{Otherwise}.
\end{array}
\right.
\end{equation}

\subsubsection{Number Optimization}
Similarly, given the transmission time, the optimal phase shift, and the amplification factor, we focus on optimizing the number of active units to minimize energy consumption. Then, problem (\ref{eq59}) can be reduced to 
\begin{equation}\label{eq69}
\begin{array}{l}
\min\limits_{\mbox{\scriptsize$\begin{array}{c} 
		N_{\rm act}
		\end{array}$}} 
\sum\limits_{k=1}^Kt_k^*(P_k^{\max}+P_k^{\max}|{h}_{{\rm r}, k}|^2N_{\rm act}(\rho_k^*)^{2}\\
~~~~~~~{+}\sigma^2N_{\rm act}(\rho_k^*)^{2}+NP_{\rm C}{+}N_{\rm act}P_{\rm DC})\\
s.t.
\tilde C_2,\tilde C_4,\bar C_5:~0\leq N_{\rm act}\leq N.\\
\end{array}
\end{equation}
It can be observed that energy consumption is an increasing function of $N_{\rm act}$, and the minimum energy consumption can be achieved at the lower bound of $N_{\rm act}$. By analyzing $\tilde C_4$, we have
\begin{equation}\label{eq70}
\begin{array}{l}
\frac{P_k^{\max}|f_k|^2((\rho_k^*-1)N_{\rm act}+N)^2}{\sigma^2|h|^2(\rho_k^*)^2N_{\rm act}+\delta^2}\geq \hat S_k\\
\overset{(c)}{\Rightarrow}\frac{4P_k^{\max}|f_k|^2(\rho_k^*-1)NN_{\rm act}}{\sigma^2|h|^2(\rho_k^*)^2N_{\rm act}+\delta^2}\geq \hat S_k\\
\Rightarrow N_{\rm act}\geq\frac{\delta^2\hat S_k}{4P_k^{\max}|f_k|^2(\rho_{k}^*-1)N-\sigma^2|h|^2(\rho_{k}^*)^2\hat S_k},
\end{array}
\end{equation}
where $(c)$ is due to $(a+b)^2\geq 4ab$. When $4P_k^{\max}|h_{{\rm r},k}|^2(\rho_{k}^*-1)N\geq\sigma^2(\rho_{k}^*)^2\hat S_k$, the lower bound of $N_{\rm act}$ can be expressed as
\begin{equation}\label{eq70-1}
\begin{array}{l}
N_{\rm act}^{\rm D}=\frac{\delta^2\hat S_k}{4P_k^{\max}|f_k|^2(\rho_{k}^*-1)N-\sigma^2|h|^2(\rho_{k}^*)^2\hat S_k}.
\end{array}
\end{equation}
Otherwise, there exists no feasible solution. Besides, based on $\tilde C_2$ and $\bar C_5$, we can obtain the upper bound of $N_{\rm act}$, which can be expressed as
\begin{equation}\label{eq70-2}
\begin{array}{l}
N_{\rm act}^{\rm U}=\min\left\{\frac{P_{\rm R}^{\max}}{( P_k^{\max}|h_{{\rm r}, k}|^2+\sigma^2)(\rho_{k}^*)^2},N\right\}.
\end{array}
\end{equation} 
By comparing (\ref{eq70-1}) and (\ref{eq70-2}), we have the following Proposition 4. The number tradeoff of active and passive units for energy consumption minimization is summarized in \textbf{Algorithm 3}.
\begin{spacing}{1.00}
	\floatname{algorithm}{Algorithm}
	\renewcommand{\algorithmicrequire}{\textbf{Input:}}
	\renewcommand{\algorithmicensure}{\textbf{Output:}}
	\begin{algorithm}[!t]
		\small
		\caption{: The Number Tradeoff of Active and Passive Units for Energy Consumption Minimization}
		\begin{algorithmic}[1]
			\State Initialize system parameters: $K$, $N$, $P_{\rm R}^{\max}$, $P_k^{\max}$, $S_k$, $\delta^2$, $\sigma^2$, $P_{\rm C}$, $P_{\rm DC}$, $B$;
			\State Calculate the phase shift $\theta_{k,n}^*$ by (\ref{eq60});
			\State Substitute the phase shift $\theta_{k,n}^*$ into problem (\ref{eq59}), 
			\Repeat
			\State Calculate the transmission time $t_k^*$ by (\ref{eq63}) for given the amplification factor and the number of active and passive units;
			\State Calculate the amplification factor $\rho_k^*$ by (\ref{eq68}) for given the transmission time and the number of active and passive units;
			\State Calculate the number of active and passive units $N_{\rm act}^*$ and $N_{\rm pas}^*$ by (\ref{eq70-3}) for given the transmission time and the amplification factor;
			\Until{Convergence}
		\end{algorithmic}
	\end{algorithm}
\end{spacing} 

\textit{\textbf{Proposition 4:}} The closed-form solution of the number of active units can be obtained when the following condition holds.
	\begin{equation}\label{eq70-3}
	N_{\rm act}^*{=}\left\{
	\begin{array}{l}
	N_{\rm act}^{\rm D}, 4P_k^{\max}|h_{{\rm r},k}|^2(\rho_{k}^*{-}1)N{\geq}\sigma^2(\rho_{k}^*)^2\hat S_k \&N_{\rm act}^{\rm D}{\leq} N_{\rm act}^{\rm U},\\
	\textrm{Infeasible},~~~~ \textrm{Otherwise}.
	\end{array}
	\right.
	\end{equation} 

\textit{\textbf{Remark 4:}} According to (\ref{eq65}) and (\ref{eq70-3}), we can observe that the number of active units decreases with the increasing $\rho_{k}^*$, which indicates that there still exists a tradeoff between amplification factor and the number of active units for energy consumption minimization problem. The number of active units is small, and approaches zero when the transmit power is high, the channel conditions are good, and the offloading task is small. However, when the offloading tasks are large, there may be offloading outages due to poor channel conditions or low transmit power. Then, the amplification factor can be increased to improve the offloading rate and thus expand the feasible region of $\tilde C_4$.

\begin{figure}
	\centering
	\includegraphics[width=1.5in]{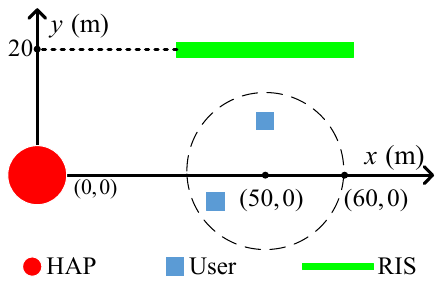} 
	\caption{The simulated scenario of the considered hybrid RIS-aided system.}
	\label{fig0-2}
\end{figure}

\begin{figure*}[htbp]\vspace{-30pt}
	\centering
	\begin{subfigure}{0.245\linewidth}
		\centering
		\includegraphics[width=1.75in]{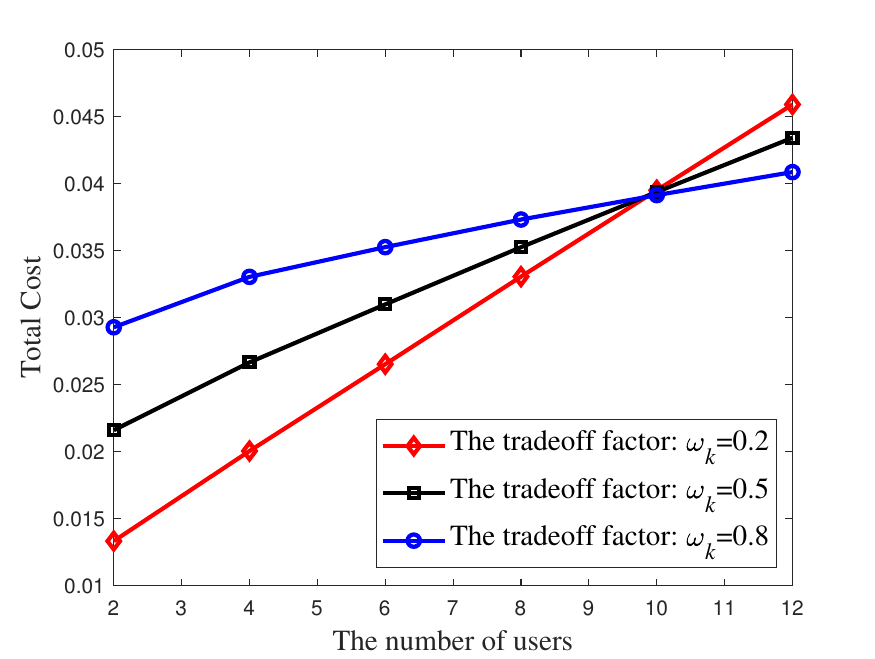}
		\caption{Total cost versus the number of users.}
		\label{chutian3}
	\end{subfigure}
	\centering
	\begin{subfigure}{0.245\linewidth}
		\centering
		\includegraphics[width=1.75in]{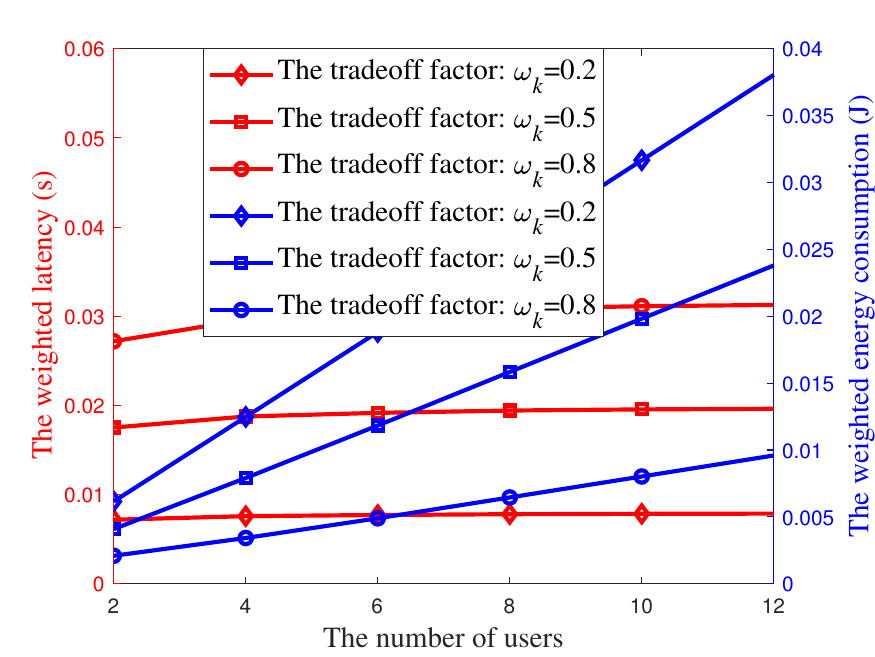}
		\caption{The weighted latency and energy consumption versus the number of users.}
		\label{chutian3}
	\end{subfigure}
	\begin{subfigure}{0.245\linewidth}
		\centering
		\includegraphics[width=1.75in]{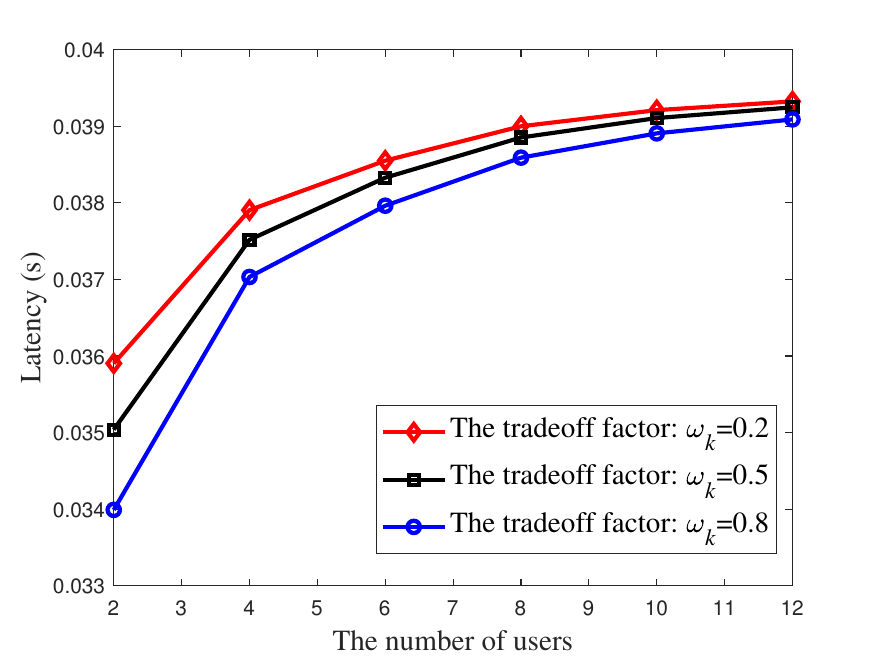}
		\caption{The latency versus the number of users.}
		\label{chutian3}
	\end{subfigure}
	\begin{subfigure}{0.245\linewidth}
		\centering
		\includegraphics[width=1.75in]{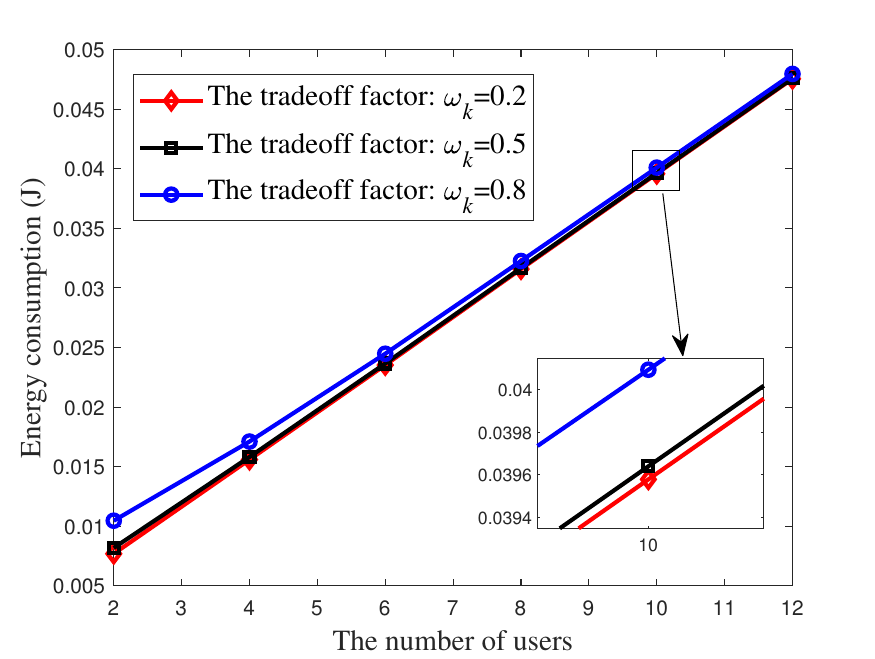}
		\caption{Energy consumption versus the number of users.}
		\label{chutian3}
	\end{subfigure}
	\caption{Tradeoff between energy consumption and the latency under different numbers of users.}
	\label{fig2}
\end{figure*}
\vspace{-10pt}
\section{Simulation Results}
In this section, the effectiveness of the proposed algorithm is evaluated by comparing it with baseline algorithms, where baseline algorithms are defined as
\begin{itemize}
	\item \textbf{Fully active RIS}\cite{i2-4}: The resource allocation algorithm for an RIS-aided
	network is performed subject to the same constraints and the objective function set as in (\ref{eq12}), but all the reflecting units are switched to the active mode.
	\item \textbf{Fully passive RIS}\cite{d1}: The resource allocation algorithm for an RIS-aided
	network is performed subject to the same constraints and the objective function set as in (\ref{eq12}), but all the reflecting units are switched to the passive mode.
	\item \textbf{Fully local computing}: The algorithm belongs to one of the binary offloading algorithms,  and users compute total tasks locally. 
	\item \textbf{Fully offloading}\cite{i2-5}: The algorithm belongs to one of the binary offloading algorithms, and users offload the total tasks to the MEC for computation.
\end{itemize}

We consider a two-dimensional coordinate setup measured in meter (m) is considered, as shown in Fig. \ref{fig0-2}, where the AP and the RIS are located at (0, 0) and (50, 20), while users are uniformly and randomly distributed in a circle centered at (50, 0) with a radius 10 m. The path-loss mode is $A=A_0(\frac{d}{d_0})^{-\beta}$, where $A_0=-30$ dB is the path-loss factor at $d_0=1$ m, $d$ is the distance between the transmitter and the receiver. The path-loss factors from the AP-RIS links, the RIS-user links, and the AP-user links are 2.6, 2.2, and 3.2, respectively. The small-scale fading follows the Rayleigh distribution\cite{i3}. Except for Fig. \ref{fig3}, the results of the other figures are obtained after 100 channel realizations. Other parameters are listed in Table I, which is similar to the setting in \cite{e2,e4}.
\begin{spacing}{1.00}
	\begin{table}[!t] \vspace{-15pt}
		\newcommand{\tabincell}[2]{\begin{tabular}{@{}#1@{}}#2\end{tabular}}
		\centering
		\scriptsize
			\caption{System Parameters.}
			\begin{tabular}{|c|c|c|c|}				
				\hline				
				\tabincell{c}{\textbf{Notation}} & \tabincell{c}{\textbf{Value}} & \tabincell{c}{\textbf{Notation}} & \tabincell{c}{\textbf{Value}}\\
				\hline
				
				\tabincell{c}{$M$} & \tabincell{c}{8} &  \tabincell{c}{$K$} & \tabincell{c}{2} \\			
				\hline
				
				\tabincell{c}{$N$} & \tabincell{c}{6} & \tabincell{c}{$ E_k^{\max}$} & \tabincell{c}{0.1 J}  \\
				\hline
				
				\tabincell{c}{$P_{\rm R}^{\max}$} & \tabincell{c}{10 dBm} & \tabincell{c}{$T_k^{\max}$} & \tabincell{c}{0.5 S}  \\
				\hline
				
				\tabincell{c}{$S_k$} & \tabincell{c}{1 Mbits} & \tabincell{c}{$C_k$} & \tabincell{c}{$4\times7^{10}$ cycles}\\
				
				\hline
				
				\tabincell{c}{$f_k^{\max}$} & \tabincell{c}{1 GHz} & \tabincell{c}{$\delta^2$} & \tabincell{c}{-80 dBm} \\

				\hline
				
				\tabincell{c}{$\sigma^2$} & \tabincell{c}{-80 dBm} & \tabincell{c}{$P_{\rm C}$} & \tabincell{c}{-10 dBm} \\
				
				\hline
				
				\tabincell{c}{$P_{\rm DC}$} & \tabincell{c}{-5 dBm}  & \tabincell{c}{$\psi$} & \tabincell{c}{1} \\			
				
				\hline
				
				\tabincell{c}{$\kappa$} & \tabincell{c}{$10^{-28}$}  & \tabincell{c}{$B$} & \tabincell{c}{1 MHz} \\
				
				\hline
				
				\tabincell{c}{$\rho_{k,n}$} & \tabincell{c}{[1,14]} & \tabincell{c}{$L_{\max}$ and $\epsilon$} & \tabincell{c}{$10^3$ and $10^{-3}$} \\
				
				\hline
				
		\end{tabular} 
	\end{table}	
\end{spacing}	
\vspace{-10pt}
\subsection{Performance Analysis of the Proposed Algorithm}
 \begin{figure}\vspace{-10pt}
	\centering
	\includegraphics[width=3in]{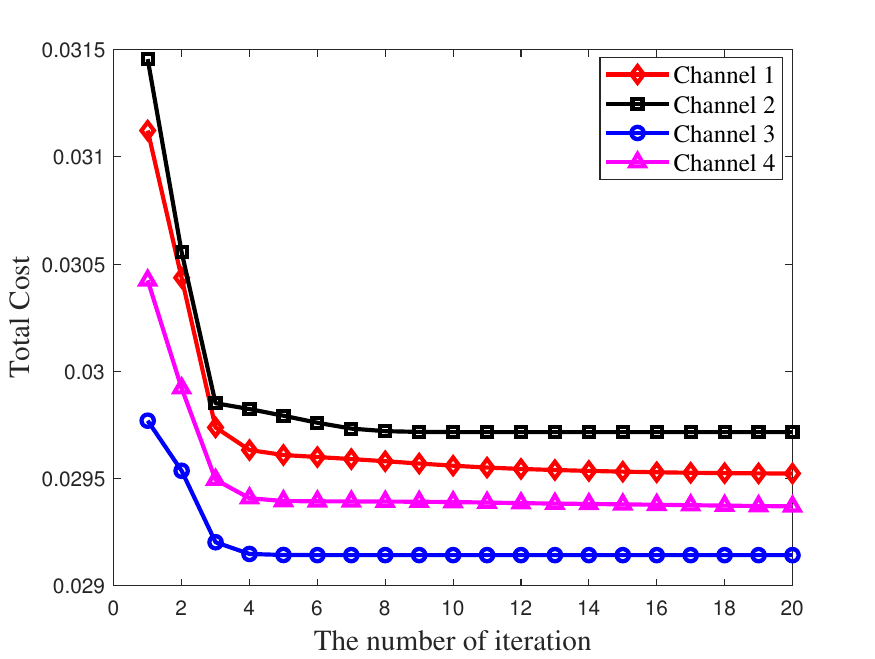}
	\caption{Convergence analysis for the proposed algorithm.}
	\label{fig3}
\end{figure}
\begin{figure}[t]\vspace{-23pt}
	\centering
	\includegraphics[width=3in]{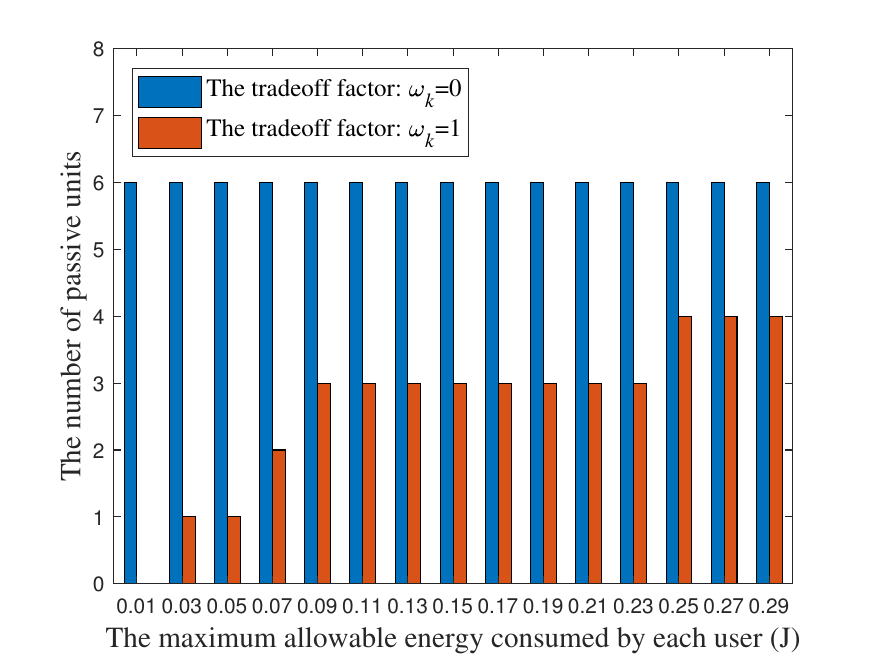}
	\caption{The number of passive units versus the maximum energy consumed by each user.}
	\label{fig4}
\end{figure}
Fig. \ref{fig2} shows the tradeoff between energy consumption and latency under different numbers of users. It can be observed from Fig. \ref{fig2}(a) that the total cost increases with the increasing number of users. The reason is that with the increasing number of users, the latency of the TDMA-based system will increase. At the same time, in order to ensure the basic computing and offloading requirement of the task for each user, the energy consumption of the system will also increase, which will lead to an increase in total cost. Besides, the total cost increases with the increase of the tradeoff factor $w_k$ when the number of users is small. However, when the number of users is large, the total cost decreases with the increasing $w_k$. This phenomenon can be explained by Fig. \ref{fig2}(b)-Fig. \ref{fig2}(d), and there are two reasons behind it. First, we can observe from Fig. \ref{fig2}(c) and Fig. \ref{fig2}(d) that the latency decreases and energy consumption increases with the increasing $w_k$. This is because larger $w_k$ indicates that the system is inclined to minimize the latency, and the proportion of energy consumption in total cost will be reduced, which will lead to an increase in the total energy consumption. However, the trend of the weighted latency and weighted energy consumption is the opposite. Second, as shown in Fig. \ref{fig2}(b), the weighted latency increases with the increasing $w_k$ when the number of users is less than 10, then the weighted latency is larger than the weighted energy consumption, and the weighted latency is dominant. Therefore, the total cost increases with the increasing $w_k$. However, as the number of users increases, energy consumption increases dramatically. The smaller $w_k$ is, the greater the weighted energy consumption is. Then, the weighted energy consumption is greater than the weighted latency, and the weighted energy consumption is dominant. Therefore, the total cost decreases with the increasing $w_k$.

Fig. \ref{fig3} shows the convergence behavior of the proposed algorithm under different channel realizations. It can be observed that the total cost gradually decreases with the increasing number of iterations. After approximately ten iterations, the total cost converges to a stable value, which reveals that the proposed algorithm has good convergence performance.

\begin{figure}[t]\vspace{-30pt}
	\centering
	\includegraphics[width=3in]{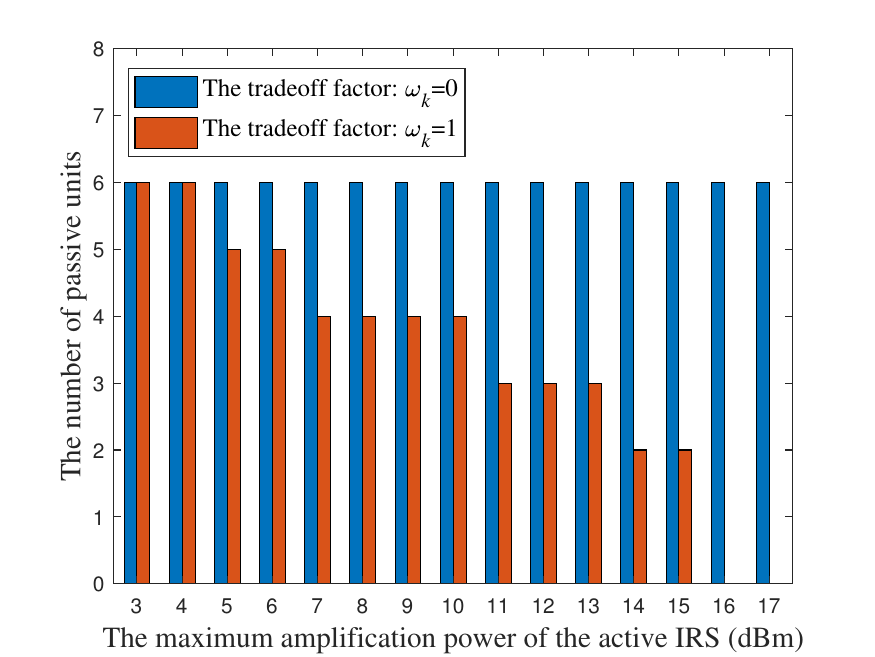}
	\caption{The number of passive units versus the maximum amplification power of the active RIS.}
	\label{fig5}
\end{figure}
\begin{figure}[t]\vspace{-20pt}
	\centering
	\includegraphics[width=3in]{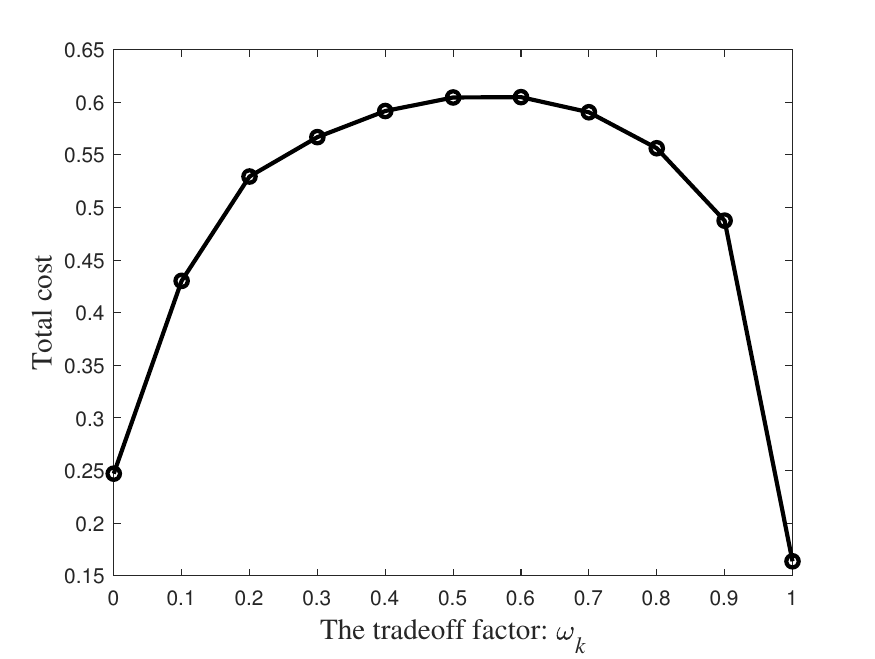}
	\caption{Total cost versus the tradeoff factor.}
	\label{fig6}
\end{figure}

\begin{figure}[t]\vspace{-30pt}
	\centering
	\includegraphics[width=3in]{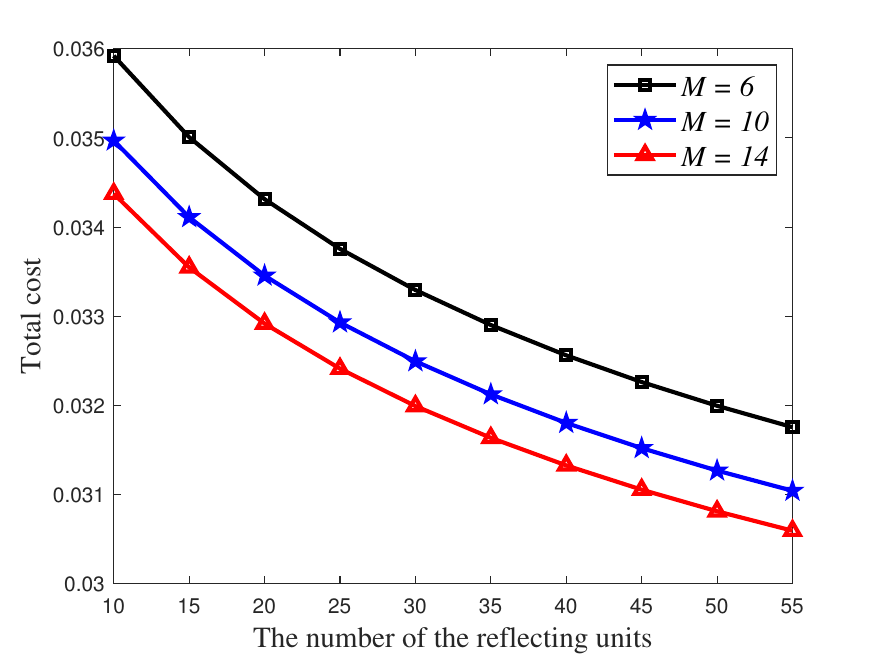}
	\caption{Total cost versus the number of the reflecting units.}
	\label{fig6-1}
\end{figure}
\begin{figure}\vspace{-15pt}
	\centering
	\includegraphics[width=3in]{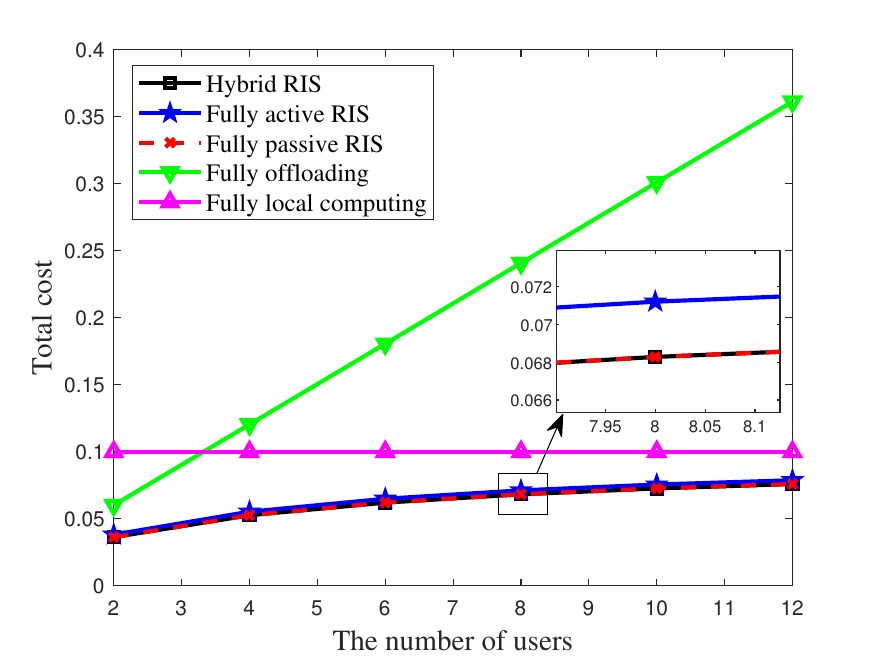} 
	\caption{Total cost versus the number of users under different algorithms.}
	\label{fig9-0}
\end{figure}

Fig. \ref{fig4} shows the number of passive units versus the maximum energy consumed by each user under different $w_k$. It can be observed that the number of passive units increases and the number of active units decreases with the increasing maximum energy consumed by each user when $w_k=1$, and the number of active units is zero when $w_k=0$. This is because when $w_k=1$, the original optimization problem is reduced to a latency minimization problem, and the system allocates more resources to minimize the latency, which means that the system should not only increase the transmit power but more reflecting units need to be switched to active mode so that the latency can be reduced. Second, the increasing maximum energy increases the feasible region of available transmit power, a higher transmit power can decrease the number of active reflecting units. Besides, the original optimization problem is reduced to an energy consumption minimization problem when $w_k=0$. In order to ensure the offloading requirements and decrease energy consumption, reflecting units will not be switched to the active mode but adopt the passive mode to offload the task when the channel condition is good.

Fig. \ref{fig5} shows the number of passive units versus the maximum amplification power of the active RIS. It can be observed that the number of passive units decreases with the increasing maximum amplification power of the active RIS when $w_k=1$ and the number of active units is always zero when $w_k=0$. The reason is that the hybrid RIS has higher power to minimize the latency when $w_k=1$, thus more reflecting units are switched to the active mode. Moreover, when $w_k=0$, although the amplified power of hybrid RIS is improved, the system aims to minimize energy consumption, and reflecting units will not be switched to the active mode when the channel condition is good.

Fig. \ref{fig6} shows the tradeoff between latency and energy consumption with respect to the tradeoff factor $w_k$. From the figure, the total cost increases with the increasing $w_k$ and then it decreases. This is because the latency decreases with $w_k$ and the energy consumption increases with $w_k$. Larger $w_k$ means that users have a higher weight on the latency and a lower weight on energy consumption. Thus, with the $w_k$ increases, the latency decreases and energy consumption increases.

Fig. \ref{fig6-1} illustrates the impact of different numbers of reflecting units on the total cost. From the figure, we can observe that as the number of reflecting units increases, the total cost gradually decreases. This is because with the increasing number of reflecting units, the system can achieve larger beamforming gains, thereby reducing the total cost. Additionally, the total cost decreases as the number of antennas at the AP increases. This is because increasing the number of receive antennas can enhance spatial diversity, thereby improving the received signal and reducing the total cost.

\subsection{Performance Comparison of the Proposed Algorithm}
Fig. \ref{fig9-0} shows the total cost versus the number of users under different algorithms. It can be observed that the total cost under all algorithms increases with the increasing number of users. This is because the increase in the number of users leads to longer offloading times and consumes more energy, resulting in an overall increase in the total cost. Meanwhile, we can observe that the performance of the proposed hybrid RIS is equivalent to that of the fully passive RIS and outperforms the fully active RIS. This is because the increase in the number of users leads to a significant rise in energy consumption, and driving the fully active RIS requires even higher energy consumption. Therefore, the reflecting units in the hybrid RIS are switched to passive mode to reduce the overall cost. Furthermore, when the number of users is large, the performance of fully local computing is better than that of fully offloading since users' local computing can be executed in parallel, resulting in a lower latency and also leading to a slow increase in total cost. On the other hand, fully offloading operates in a TDMA manner, and its disadvantage becomes more apparent as the number of users increases.
\begin{figure*}[htbp]\vspace{-30pt}
	\centering
	\begin{subfigure}{0.49\linewidth}
		\centering
		\includegraphics[width=3in]{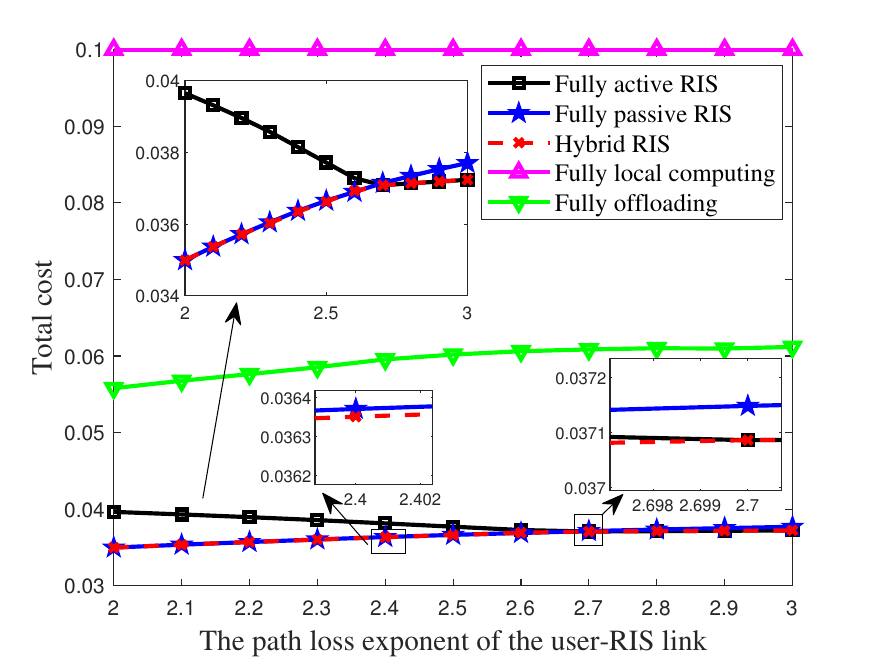}
		\caption{Total cost versus the path loss exponent.}
		\label{chutian3}
	\end{subfigure}
	\centering
	\begin{subfigure}{0.49\linewidth}
		\centering
		\includegraphics[width=3in]{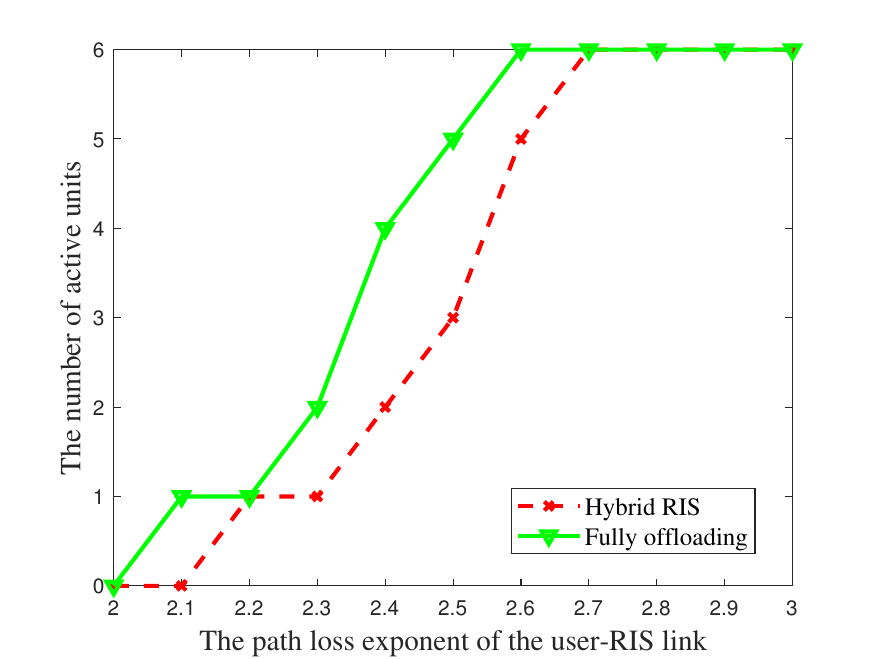}
		\caption{The number of active units versus the path loss exponent.}
		\label{chutian3}
	\end{subfigure}
	\caption{Impact of the path loss exponent on the total cost and the number of active units.}
	\label{fig8}
\end{figure*}
\begin{figure*}[htbp]\vspace{-15pt}
	\centering
	\begin{subfigure}{0.49\linewidth}
		\centering
		\includegraphics[width=3in]{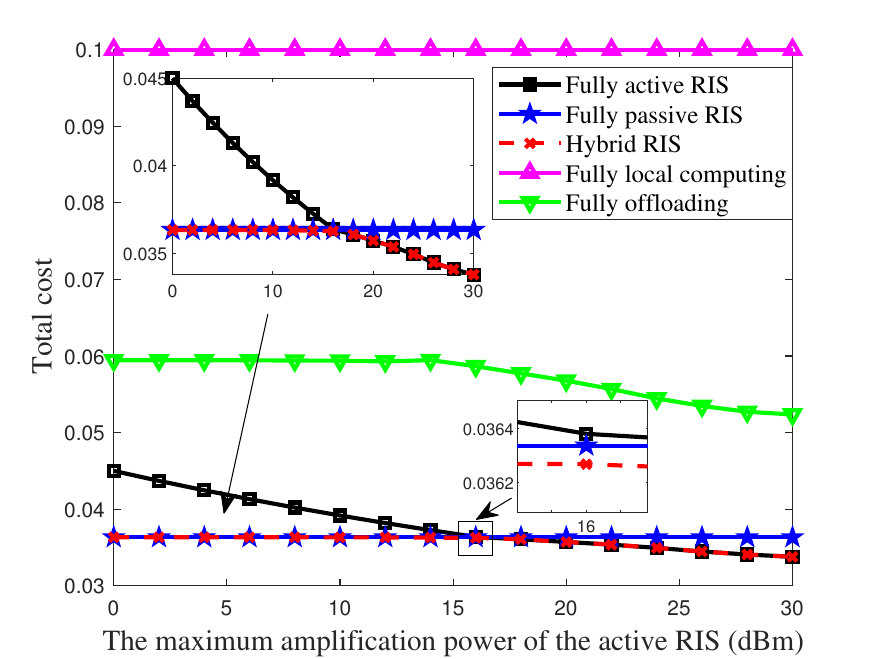}
		\caption{Total cost versus the maximum amplification power.}
		\label{chutian3}
	\end{subfigure}
	\centering
	\begin{subfigure}{0.49\linewidth}
		\centering
		\includegraphics[width=3in]{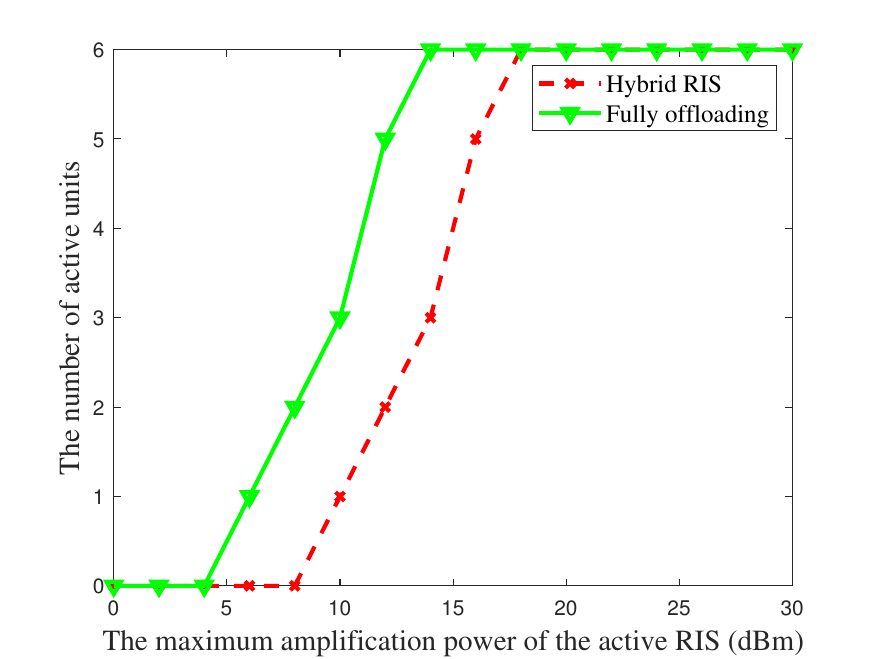}
		\caption{The number of active units versus the maximum amplification power.}
		\label{chutian3}
	\end{subfigure}
	\caption{Impact of the maximum amplification power on the total cost and the number of active units.}
	\label{fig9}\vspace{-10pt}
\end{figure*}

Fig. \ref{fig8} shows the total cost and the number of active units versus the path loss exponent of the user-RIS link under different algorithms. From Fig. \ref{fig8}(a), we can see that the total cost under all algorithms increases with the increasing path loss exponent except for the fully active RIS algorithm and the fully local computing algorithm. This is because the larger the path loss exponent is, the worse the channel condition is, which directly reduces the offloading rates of users and increases the offloading latency. Then the system has to increase the transmit power and more reflecting units are switched to the active mode to overcome the above problems, which also increases the energy consumption. Total cost under the fully active RIS algorithm first decreases and then increases with the increasing path loss exponent. From Fig. \ref{fig8}(b), the number of active units is zero when the path loss exponent is small, thus the SINR under this algorithm exists amplification noise, which increases the latency and indicates that the total cost is higher than that under the fully passive RIS algorithm. Then, the total cost will decrease since users allocate more transmit power to decrease the latency until the transmit power reaches saturation and then the number of active units reaches the maximum. The total cost will increase when the path loss exponent further increases. Besides, the total cost under the fully local computing algorithm remains unchanged since the resource allocation under this algorithm is independent of the path loss exponent. Furthermore, the hybrid RIS algorithm first adopts the fully-passive mode and then gradually switches to the fully-active mode to ensure low cost, as shown in (\ref{eq57}), which verifies the superiority of the hybrid RIS algorithm.

Fig. \ref{fig9} shows the total cost and the number of active units versus the maximum amplification power under different algorithms. It can be observed that the total cost under all algorithms decreases with the increasing maximum amplification power expected for the fully local computing algorithm since the feasible region of amplification power is expanded and the RIS is not adopted by the fully local computing algorithm. From Fig. \ref{fig9}(a), the fully local computing algorithm has the highest cost, followed by the fully offloading algorithm, next is the partial offloading algorithm, i.e., the fully active RIS algorithm, the fully passive algorithm, and the hybrid RIS algorithm, which reflects the advantages of task offloading and the effectiveness of partial offloading. In addition, all reflecting units under the hybrid RIS algorithm are switched to the passive mode when amplification power is low, and then gradually are switched to the active mode with the increasing amplification power threshold. Compared with the fully active RIS algorithm and the fully passive RIS algorithm, the hybrid RIS algorithm has a lower cost, which reveals that the hybrid RIS algorithm has better flexibility. Furthermore, from Fig. \ref{fig9}(b), the number of active units under the hybrid RIS algorithm is less than that under the fully offloading algorithm in a certain feasible region since part of the tasks under the hybrid RIS algorithm is computed locally, while the fully offloading algorithm requires more number of active units to realize task offloading.

\begin{figure*}[htbp]\vspace{-30pt}
	\centering
	\begin{subfigure}{0.49\linewidth}
		\centering
		\includegraphics[width=3in]{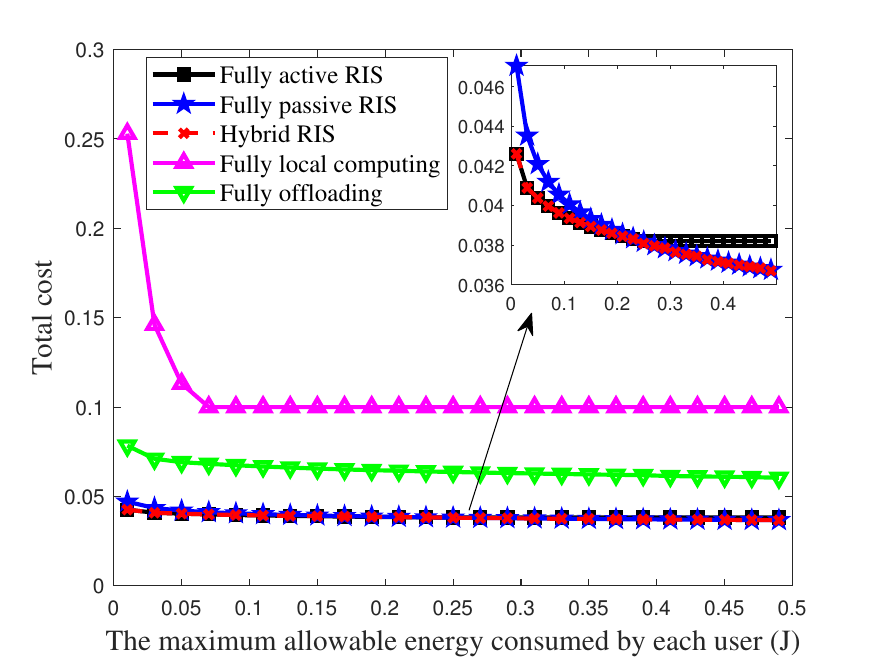}
		\caption{Total cost versus the maximum energy consumption.}
		\label{fig10a}
	\end{subfigure}
	\centering
	\begin{subfigure}{0.49\linewidth}
		\centering
		\includegraphics[width=3in]{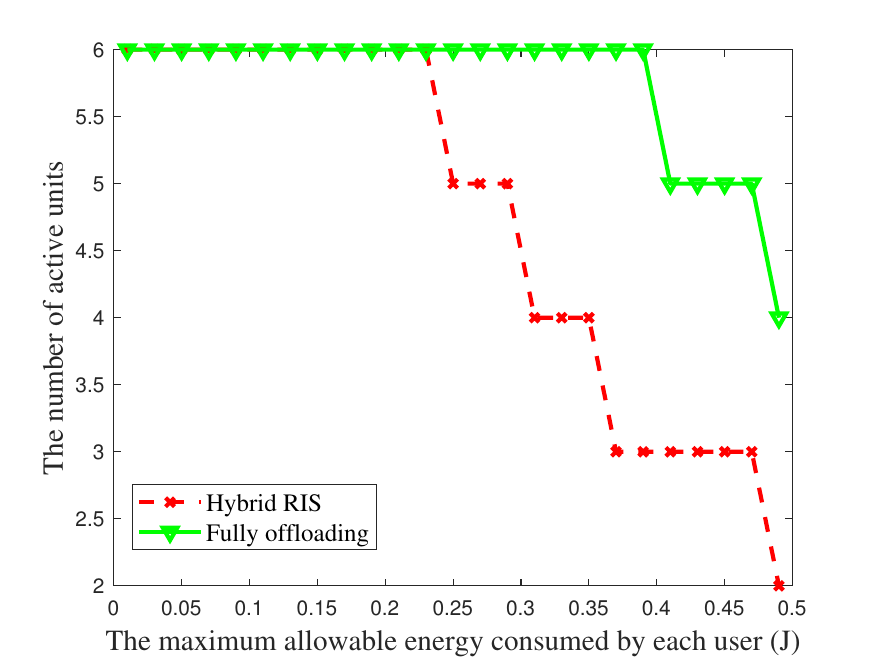}
		\caption{The number of active units versus the maximum energy consumption.}
		\label{fig10b}
	\end{subfigure}
	\caption{Impact of the maximum energy consumption on the total cost and the number of active units.}
	\label{fig10}
\end{figure*}
Fig. \ref{fig10} shows the total cost and the number of active units versus the maximum energy consumption under different algorithms. It can be observed that the total cost under all algorithms decreases with the increasing maximum energy consumed by each user since users have more energy to perform local computing and have higher transmit power to perform task offloading. It is worth noting that the total cost under the fully local computing algorithm decreases first and then keeps the same with the increasing maximum energy threshold because although the maximum energy threshold increases, local computing is still limited by the maximum computing capacity. Besides, when the maximum energy threshold is small, the fully active algorithm has a lower cost since the maximum energy threshold limits the user's transmit power, making the offloading rate under the fully passive RIS algorithm low. Then, the fully active RIS algorithm can directly improve the offloading rate by amplifying the received signal and the number of active units is large from Fig. \ref{fig10}(b). Furthermore, users can allocate a higher transmit power to improve the offloading rate and the number of active units decreases gradually from Fig. \ref{fig10}(b) when the maximum energy threshold becomes large. Besides, the total cost under the fully passive RIS algorithm is lower than that under the fully active RIS algorithm when the maximum energy threshold is large since the energy consumption under the fully passive RIS algorithm is low and the amplification noise caused by the fully active RIS algorithm increases the latency. Furthermore, the proposed hybrid RIS algorithm outperforms other algorithms, which verifies the effectiveness of the hybrid RIS algorithm. 
\begin{figure}[!t]\vspace{-20pt}
	\centering
	\includegraphics[width=3in]{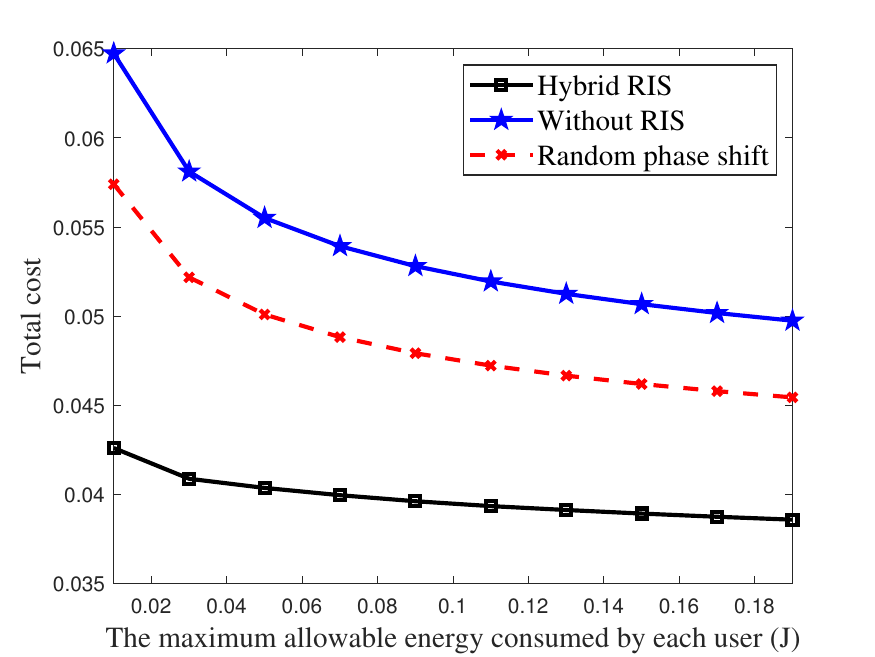}
	\caption{Total cost versus the maximum energy consumed by each user.}
	\label{fig12}
\end{figure}

Fig. \ref{fig12} evaluates the impact of the maximum energy consumed by each user on total cost under different general algorithms, such as without RIS algorithm and random phase shift algorithm. From the figure, we can observe that the total cost under the proposed hybrid RIS algorithm is lower than that of the without RIS algorithm and the random phase shift algorithm. This is an anticipated result because the deployment of the RIS provides additional reflecting links to enhance the offloading capability, which is crucial for MEC offloading strategies. Undoubtedly, the random phase shift algorithm is typically employed to reduce the feedback overhead, but its performance is indeed inferior.
\vspace{-10pt}
\section{Conclusion}
This paper proposes a novel hybrid active-passive RIS and analyzes the impact of the mode switching of reflecting units on the MEC-aided system. In particular, a problem is formulated with the objective of minimizing total cost in terms of the latency and energy consumption while satisfying constraints on the maximum energy of users, the maximum power of the RIS, the minimum computation tasks, the transmission time, the offloading ratio factor, the mode-switching factor, the unit moduli of passive units, and the computation ability. To solve this problem, we develop an AO-based iterative algorithm adopting the variable substitution method, the SVD method, and the SCA method. Moreover, we reduce the original problem to a latency minimization problem and an energy consumption minimization problem, and explore the tradeoff between the number of active and passive units. Simulation results demonstrate that the proposed algorithm outperforms baseline algorithms and can flexibly switch between active and passive modes.
\vspace{-10pt}
\appendices
\section{}
Given a maximum power threshold of the RIS $P_{\rm R}^{\max}$, we can observe that constraint $\tilde C_2$ is active when the offloading rate is maximized. Then, we have
\begin{equation}\label{eq71}
\begin{array}{l}
 P_k^{\max}|{h}_{{\rm r}, k}|^2N_{\rm act}\rho_{k}^2+\sigma^2N_{\rm act}\rho_{k}^2= P_{\rm R}^{\max}.\\
\end{array}
\end{equation}
From the formula (\ref{eq71}), we can know that $N_{\rm act}\rho_{k}^2$ is a constant. Then, the offloading rate increases with the increasing $\rho_{k}$. Thus, we can maximize offloading rate by maximizing the numerator of the SINR expression, i.e.,
\begin{equation}\label{eq72}
\begin{array}{l}
P_k^{\max}|f_k|^2\left(N_{\rm act}\rho_{k}+N-N_{\rm act}\right)^2.
\end{array}
\end{equation}
We can observe that the maximum offloading rate can be achieved  when $\rho_k$ reaches its maximum. Thus, the amplification factor can be obtained by the following transformation
\begin{equation}\label{eq73}
\begin{array}{l}
P_k^{\max}|{h}_{{\rm r}, k}|^2N_{\rm act}\rho_{k}^2+\sigma^2N_{\rm act}\rho_{k}^2= P_{\rm R}^{\max}
\\
\Rightarrow \rho_k= \sqrt\frac{P_{\rm R}^{\max}}{( P_k^{\max}|h_{{\rm r},k}|^2+\sigma^2)N_{\rm act}}.
\end{array}
\end{equation}

The proof is completed.

\vspace{-10pt}
\section{}
Define the objective function as $f(\sqrt{N_{\rm act}})=A_{1,k}(A_{2,k}\sqrt{N_{\rm act}}-N_{\rm act}+N)^2$, where
\begin{equation}\label{eq74}
\begin{array}{l}
A_{1,k}=\frac{P_k^{\max}|f_k|^2}{\frac{\sigma^2|h|^2P_{\rm R}^{\max}}{ P_k^{\max}|h_{{\rm r},k}|^2+\sigma^2}+\delta^2},A_{2,k}=\sqrt{\frac{P_{\rm R}^{\max}}{ P_k^{\max}|h_{{\rm r},k}|^2+\sigma^2}}.
\end{array}
\end{equation}
We take the derivative of $f(\sqrt{N_{\rm act}})$ with respect to $\sqrt{N_{\rm act}}$ and we have
\begin{equation}\label{eq75}
\begin{array}{l}
\frac{\partial f(\sqrt{N_{\rm act}})}{\partial\sqrt{N_{\rm act}}}{=}2A_{1,k}(A_{2,k}\sqrt{N_{\rm act}}{-}N_{\rm act}{+}N)(A_{2,k}{-}2\sqrt{N_{\rm act}}).
\end{array}
\end{equation}
By setting (\ref{eq75}) to $0$, we can obtain three roots, i.e.,
\begin{equation}\label{eq76}
\begin{array}{l}
x_1=\frac{\sqrt{\frac{P_{\rm R}^{\max}}{ P_k^{\max}|h_{{\rm r},k}|^2+\sigma^2}}-\sqrt{\frac{P_{\rm R}^{\max}}{ P_k^{\max}|h_{{\rm r},k}|^2+\sigma^2}+4N}}{2}<0,
\end{array}
\end{equation}
\begin{equation}\label{eq77}
\begin{array}{l}
x_2=\frac{\sqrt{\frac{P_{\rm R}^{\max}}{ P_k^{\max}|h_{{\rm r},k}|^2+\sigma^2}}}{2}>0,
\end{array}
\end{equation}
\begin{equation}\label{eq78}
\begin{array}{l}
x_3=\frac{\sqrt{\frac{P_{\rm R}^{\max}}{ P_k^{\max}|h_{{\rm r},k}|^2+\sigma^2}}+\sqrt{\frac{P_{\rm R}^{\max}}{ P_k^{\max}|h_{{\rm r},k}|^2+\sigma^2}+4N}}{2}>0.
\end{array}
\end{equation}
It is noted that $x_1$ should be abandoned due to $x_1<0$. Meanwhile, we have $x_2<x_3$ and furthermore the following Table II.
\begin{table}[!h]
	\newcommand{\tabincell}[2]{\begin{tabular}{@{}#1@{}}#2\end{tabular}}
	\centering
	\footnotesize
	\caption{}
		\begin{tabular}{|c|c|c|c|c|c|}
			\hline
			\tabincell{c}{} & \tabincell{c}{$[0,x_2]$} & \tabincell{c}{$x_2$} & \tabincell{c}{$[x_2,x_3]$} &
			\tabincell{c}{$x_3$} & \tabincell{c}{$[x_3,+\infty)$}\\
			\hline
			\tabincell{c}{$\frac{\partial f(\sqrt{N_{\rm act}})}{\partial\sqrt{N_{\rm act}}}$} & 
			\tabincell{c}{$+$} & \tabincell{c}{$0$} & \tabincell{c}{$-$}
			& \tabincell{c}{$0$}
			& \tabincell{c}{$+$}\\
			\hline
			\tabincell{c}{$f(\sqrt{N_{\rm act}})$} & \tabincell{c}{$\uparrow$} & \tabincell{c}{Maximum} & \tabincell{c}{$\downarrow$}
			& \tabincell{c}{Minimum}
			& \tabincell{c}{$\uparrow$}\\
			\hline
	\end{tabular}   
\end{table}
From Table II, we can know that the maximum rate can be achieved when $N_{\rm act} = x_2^2$. Then, there exist three cases: 
\begin{itemize}
	\item Case I: When $\sqrt{N}\leq\frac{\sqrt{\frac{P_{\rm R}^{\max}}{ P_k^{\max}|h_{{\rm r},k}|^2+\sigma^2}}}{2}=x_2$ holds, i.e., $h_{{\rm r},k}\leq\sqrt{\frac{P_{\rm R}^{\max}}{4N P_k^{\max}}-\frac{\sigma^2}{ P_k^{\max}}}$. Then, we have $N_{\rm act}=N$ and $N_{\rm pas}=0$.
	\item Case II: When $x_2=\frac{\sqrt{\frac{P_{\rm R}^{\max}}{ P_k^{\max}|h_{{\rm r},k}|^2+\sigma^2}}}{2}<\sqrt{N}\leq\frac{\sqrt{\frac{P_{\rm R}^{\max}}{ P_k^{\max}|h_{{\rm r},k}|^2+\sigma^2}}+\sqrt{\frac{P_{\rm R}^{\max}}{ P_k^{\max}|h_{{\rm r},k}|^2+\sigma^2}+4N}}{2}$ holds, we have $N_{\rm act}=\frac{P_{\rm R}^{\max}}{4( P_k^{\max}|h_{{\rm r},k}|^2+\sigma^2)}$ and $N_{\rm pas}=N-\frac{P_{\rm R}^{\max}}{4( P_k^{\max}|h_{{\rm r},k}|^2+\sigma^2)}$.
	\item Case III: When $\sqrt{N}>\frac{\sqrt{\frac{P_{\rm R}^{\max}}{ P_k^{\max}|h_{{\rm r},k}|^2+\sigma^2}}+\sqrt{\frac{P_{\rm R}^{\max}}{ P_k^{\max}|h_{{\rm r},k}|^2+\sigma^2}+4N}}{2}$ holds, we have
	\begin{equation}
		\begin{array}{l}\label{eq79}
		\!\!\!\!\!f(x_2){-}f(\sqrt{N}){=}A_{1,k}(\frac{A_{2,k}^2}{4}{+}N{-}A_{2,k}\sqrt{N})(\frac{A_{2,k}^2}{4}\\
		\!\!\!\!\!{+}N{+}A_{2,k}\sqrt{N}){=}A_{1,k}(\frac{A_{2,k}}{2}{-}\sqrt{N})^2(\frac{A_{2,k}^2}{4}{+}N\\
		\!\!\!\!\!+A_{2,k}\sqrt{N})\geq 0.
		\end{array}
	\end{equation}	
	Then, we have
	 $N_{\rm act}=\frac{P_{\rm R}^{\max}}{4( P_k^{\max}|h_{{\rm r},k}|^2+\sigma^2)}$ and $N_{\rm pas}=N-\frac{P_{\rm R}^{\max}}{4( P_k^{\max}|h_{{\rm r},k}|^2+\sigma^2)}$.
\end{itemize}

The proof is completed.
\vspace{-10pt}
\section{}
Given the mode-switching factor, the amplification factor, and the transmission time, the minimum energy consumption can be achieved when the offloading rate is equal to the input data size, i.e., $t_kR_k=S_k$. Then, the objective function can be rewritten as
\begin{equation}\label{eq80}
\begin{array}{l}
\sum\limits_{k=1}^Kt_k(P_k^{\max}+P_k^{\rm act}+P_k^{\rm pas})=\sum\limits_{k=1}^K\frac{S_k(P_k^{\max}+P_k^{\rm act}+P_k^{\rm pas})}{R_k}=\\
\sum\limits_{k=1}^K\frac{S_k(P_k^{\max}\sum\limits_{n=1}^N(\alpha_{k,n}\rho_{k,n}^{\alpha_{k,n}}|{h}_{{\rm r}, k}^n|)^2{+}\sigma^2\sum\limits_{n=1}^N(\alpha_{k,n}\rho_{k,n}^{\alpha_{k,n}})^2{+}A_k)}{B\log_2(1+\frac{P_k^{\max}|\boldsymbol{\rm h}^H\boldsymbol{\rm \Lambda}_k\boldsymbol{\rm \Theta}_k\boldsymbol{\rm h}_{{\rm r}, k}|^2}{\sigma^2\sum\limits_{n=1}^N(\alpha_{k,n}\rho_{k,n}^{\alpha_{k,n}}|h_n|)^2{+}\delta^2})},
\end{array}
\end{equation}
where $A_k=P_k^{\max}{+}NP_{\rm C}{+}N_{\rm act}P_{\rm DC}$. Then, the minimum energy consumption can be obtained when $|\boldsymbol{\rm h}^H\boldsymbol{\rm \Lambda}_k\boldsymbol{\rm \Theta}_k\boldsymbol{\rm h}_{{\rm r}, k}|^2$ is maximized.
Thus, the optimal phase shift can be denoted as $\theta_{k,n}^*={\rm arg}([\boldsymbol{\rm h}]_n)-{\rm arg}([\boldsymbol{\rm h}_{{\rm r},k}]_n)$. 

The proof is completed.
\vspace{-10pt}
\section{}
The lower bound of $\rho_k$ can be obtained by analyzing the feasible region of $\tilde C_4$, i.e.
\begin{equation}\label{eq81}
\begin{array}{l}
\frac{P_k^{\max}|f_k|^2(N_{\rm act}\rho_k+N-N_{\rm act})^2}{\sigma^2|h|^2N_{\rm act}\rho_k^2+\delta^2}\geq \hat S_k\\
\overset{(b)}{\Rightarrow}\frac{4P_k^{\max}|f_k|^2N_{\rm act}(N-N_{\rm act})\rho_k}{\sigma^2|h|^2N_{\rm act}\rho_k^2+\delta^2}\geq \hat S_k\\
\Rightarrow-\sigma^2|h|^2N_{\rm act}\hat S_k\rho_k^2\\
+4P_k^{\max}|f_k|^2N_{\rm act}(N-N_{\rm act})\rho_k-\delta^2\hat S_k\geq 0,
\end{array}
\end{equation}
where $(b)$ is due to $(a+b)^2\geq 4ab$. Then, defining $b_k=4P_k^{\max}|f_k|^2N_{\rm act}(N-N_{\rm act})$, when $b_k^2\geq4\sigma^2\delta^2|h|^2N_{\rm act}\hat S_k^2$, we can obtain two roots, i.e.,
\begin{equation}\label{eq82}
\begin{array}{l}
r_1{=}\frac{-b_k{+}\sqrt{b_k^2{-}4\sigma^2\delta^2|h|^2N_{\rm act}\hat S_k^2}}{-2\sigma^2|h|^2N_{\rm act}\hat S_k},
\end{array}
\end{equation}
\begin{equation}\label{eq83}
\begin{array}{l}
r_2{=}\frac{-b_k{-}\sqrt{b_k^2{-}4\sigma^2\delta^2|h|^2N_{\rm act}\hat S_k^2}}{-2\sigma^2|h|^2N_{\rm act}\hat S_k},
\end{array}
\end{equation}
where $0<r_1<r_2$. Then, the inequation in (\ref{eq81}) always holds when $r_1\leq\rho_{k}\leq r_2$ holds. Thus, we have $\rho_{k}^*=\max\{r_1,\rho^{\min}\}$ when $\rho^{\min}\leq r_2$. Otherwise there exists no feasible solution.

The proof is completed.

\end{document}